\documentclass[aps,prx,twocolumn,superscriptaddress]{revtex4-2}

\usepackage[mathcal]{euscript}
\usepackage{amsmath}
\usepackage{textcomp, gensymb}
\usepackage{amssymb}
\usepackage[dvipsnames]{xcolor}
\usepackage{graphicx}

\usepackage[left]{lineno} 

\usepackage[version=3]{mhchem} 

\usepackage{xr}
\usepackage{hyperref}
\usepackage{lineno}

\newcommand{\wphi}{\widetilde{\varphi}}

\begin{document}
	\title{Dissipationless transport signature of topological nodal lines}
	
	\author{Arthur Veyrat}
	\affiliation{Leibniz Institute for Solid State and Materials Research (IFW Dresden), Helmholtzstraße 20, D-01069 Dresden, Germany}
	\affiliation{Würzburg-Dresden Cluster of Excellence ct.qmat, Dresden, Germany}
	\affiliation{Laboratoire de Physique des Solides (LPS Orsay), 510 Rue André Rivière, 91400 Orsay, France}
	
	\author{Klaus Koepernik}
	\affiliation{Leibniz Institute for Solid State and Materials Research (IFW Dresden), Helmholtzstraße 20, D-01069 Dresden, Germany}
	\affiliation{Würzburg-Dresden Cluster of Excellence ct.qmat, Dresden, Germany}
	
	\author{Louis Veyrat}
	\affiliation{Leibniz Institute for Solid State and Materials Research (IFW Dresden), Helmholtzstraße 20, D-01069 Dresden, Germany}
	\affiliation{Würzburg-Dresden Cluster of Excellence ct.qmat, Dresden, Germany}
	\affiliation{CNRS, Laboratoire National des Champs Magnétiques Intenses, Université Grenoble-Alpes, Université Toulouse 3, INSA-Toulouse, EMFL, 31400 Toulouse, France}

	\author{Grigory Shipunov}
	\author{Saicharan Aswartham}
	\author{Jiang Qu}
	\author{Ankit Kumar}
	\affiliation{Leibniz Institute for Solid State and Materials Research (IFW Dresden), Helmholtzstraße 20, D-01069 Dresden, Germany}
	\affiliation{Würzburg-Dresden Cluster of Excellence ct.qmat, Dresden, Germany}
	
	\author{Michele Ceccardi}
	\affiliation{Department of Physics, University of Genoa, 16146 Genoa, Italy}
	\affiliation{CNR-SPIN Institute, 16152 Genoa, Italy }
	
	\author{Federico Caglieris}
	\affiliation{CNR-SPIN Institute, 16152 Genoa, Italy }
	
	\author{Nicolás Pérez Rodríguez}
	\affiliation{Leibniz Institute for Solid State and Materials Research (IFW Dresden), Helmholtzstraße 20, D-01069 Dresden, Germany}
	\affiliation{Würzburg-Dresden Cluster of Excellence ct.qmat, Dresden, Germany}
	
	\author{Romain Giraud}
	\affiliation{Leibniz Institute for Solid State and Materials Research (IFW Dresden), Helmholtzstraße 20, D-01069 Dresden, Germany}
	\affiliation{Würzburg-Dresden Cluster of Excellence ct.qmat, Dresden, Germany}
	\affiliation{Université Grenoble Alpes, CNRS, CEA, Grenoble-INP, Spintec, F-38000 Grenoble, France}
	
	\author{Bernd Büchner}
	\author{Jeroen van den Brink}
	\affiliation{Leibniz Institute for Solid State and Materials Research (IFW Dresden), Helmholtzstraße 20, D-01069 Dresden, Germany}
	\affiliation{Würzburg-Dresden Cluster of Excellence ct.qmat, Dresden, Germany}
	\affiliation{Department of Physics, TU Dresden, D-01062 Dresden, Germany}
	
	\author{Carmine Ortix}
	\email{cortix@unisa.it}
	\affiliation{Dipartimento di Fisica ``E. R. Caianiello", Universit\'a di Salerno, IT-84084 Fisciano (SA), Italy}
	
	\author{Joseph Dufouleur}
	\email{j.dufouleur@ifw-dresden.de}
	\affiliation{Leibniz Institute for Solid State and Materials Research (IFW Dresden), Helmholtzstraße 20, D-01069 Dresden, Germany}
	\affiliation{Würzburg-Dresden Cluster of Excellence ct.qmat, Dresden, Germany}
	\affiliation{Center for Transport and Devices, TU Dresden, D-01069 Dresden, Germany}

	\maketitle
	\textbf{Topological materials, such as topological insulators or semimetals, usually not only reveal the nontrivial properties of their electronic wavefunctions through the appearance of stable boundary modes~\cite{Hasan2010}, but also through very specific electromagnetic responses~\cite{Qi2011}. The anisotropic longitudinal magnetoresistance of Weyl semimetals~\cite{Nandy2017}, for instance, carries the signature of the chiral anomaly of Weyl fermions. 
		However for topological nodal line (TNL) semimetals~\cite{Burkov2011,Fang2015} --  materials where the valence and conduction bands cross each other on one-dimensional curves in the three-dimensional Brillouin zone -- such a characteristic has been lacking. 
		Here we report the discovery of a peculiar charge transport effect generated by TNLs: a dissipationless transverse signal in the presence of coplanar electric and magnetic fields~\cite{Battilomo2021,Cullen2021}, which originates from a Zeeman-induced conversion of TNLs into Weyl nodes under infinitesimally small magnetic fields.
		We evidence this dissipationless topological response in trigonal \ce{PtBi2} persisting up to room temperature, and unveil the extensive TNLs in the band structure of this non-magnetic material.
		These findings provide a new pathway to engineer Weyl nodes by arbitrary small magnetic fields and reveal that 
		bulk topological nodal lines 
		can exhibit non-dissipative transport properties.}


	The electronic band structure of a bulk material can feature isolated degeneracy points where electronic states with different spin, orbital or sublattice quantum number possess same energy and crystalline momentum ${\bf k}$.
	In materials lacking either inversion or time-reversal symmetry, such degeneracies can be simply twofold while occurring at generic points in the three-dimensional Brillouin zone. The electronic bands in the vicinity of the nodes then generally resembles the energetic dispersion of massless relativistic particles governed by the Weyl equation~\cite{Armitage2018}. 
	Weyl nodes  represent monopoles of the Berry flux and are thus characterized by a well-defined topological charge. This non-trivial bulk topology is manifested in a very specific spectroscopic signature: the presence of surface Fermi arcs connecting Weyl points with opposite chirality~\cite{Wan2011,Kuibarov2024}. The characteristic electromagnetic response of Weyl quasiparticles is instead connected to their chiral anomaly~\cite{Nielsen1983,son2013}. 
	This causes a strong in-plane anisotropic magnetoconductivity that can be directly probed through measurements of the planar Hall effect (PHE)~\cite{Burkov2017,Nandy2017,Kumar2018}: the appearance, in the presence of coplanar electric and magnetic fields, of a transverse voltage with $\pi$-periodic angular dependence. Weyl quasiparticles may also be evidenced in transport experiments through other effects, such as the unconventional Hall effect\cite{Ge2020}.
	
	Point-group symmetries can also stabilize twofold degenerate closed lines in the three-dimensional Brillouin zone. When appearing at mirror invariant planes such nodal lines are characterized by a bulk $\mathbb{Z}_2$ topological invariant~\cite{Fang2015,Sun2018}. Although they are often accompanied by ``drumhead" surface states~\cite{Huang2016,Bian2016a,Chan2016}, topological nodal lines (TNLs) lack a genuine bulk-boundary correspondence: the relevant surfaces naturally break the protecting mirror symmetry~\cite{Bian2016}. Additionally, some electromagnetic responses characteristic of TNL-semimetals have been identified, but only in specific cases, such as for instance in the quantum limit~\cite{Wang2023}, making the physical consequences of the bulk topology completely hidden.
	Here, we unveil a peculiar charge transport effect associated with mirror symmetry-protected TNLs in trigonal crystals: an anomalous planar Hall effect (APHE) that is odd in magnetic field, does not contribute to the dissipated power\cite{Battilomo2021}, and is measurable in the linear transport regime. We identify trigonal-\ce{PtBi2} as an ideal material platform because of the presence of a large number of TNLs on its three vertical mirror planes that makes the anomalous planar Hall effect particularly robust and surviving up to room temperature. 
	
	\section*{Topological nodal line conversion in magnetic field}
	
	\begin{figure*}[t]
		\centering
		\includegraphics[width=1.\textwidth]{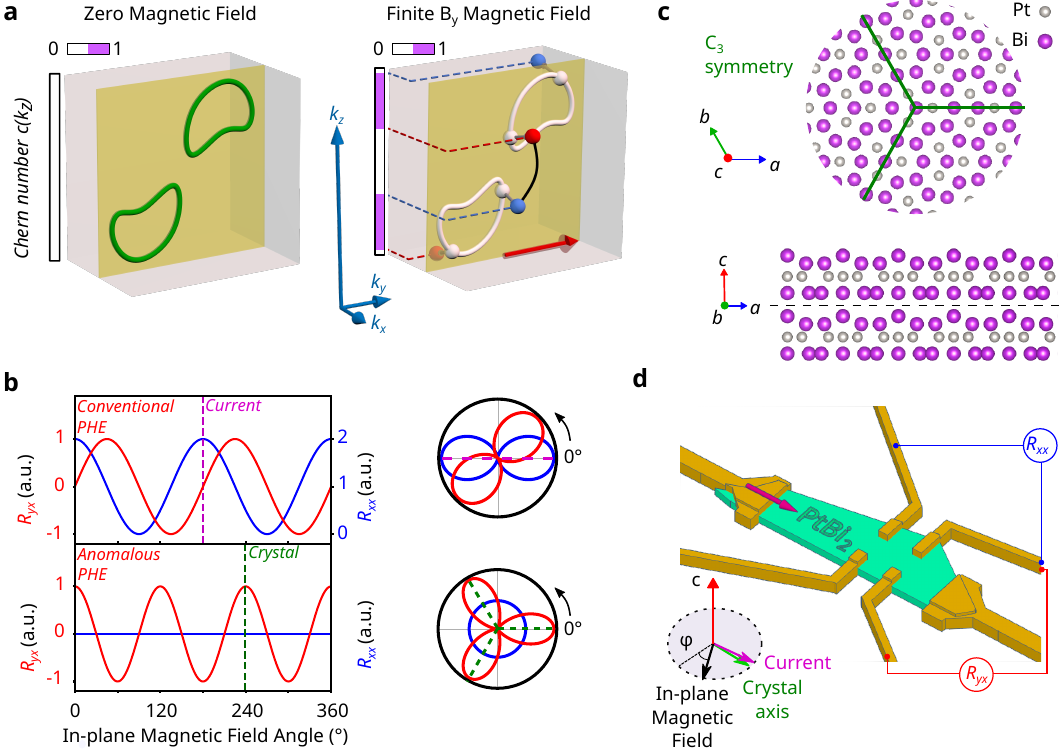}
		\caption{\textbf{Anomalous planar Hall effects in \ce{PtBi2}.} 
			\textbf{a}: Generation of an anomalous Hall conductance in nodal line semimetal systems. In a nodal line semimetal, the nodal lines do not contribute to the Chern number at zero magnetic field (left panel). Under a finite external magnetic field (right panel, red arrow), the nodal lines split into pairs of Weyl nodes of opposite chiralities. These pairs can appear anywhere on the nodal lines, including with a significant $k_z$ separation. This leads to potential large $k_z$ ranges of non-zero $c(k_z)$, inducing a large AHC at finite field. 
			\textbf{b}: Typical angular dependence of the conventional (top panel) and anomalous (bottom panel) planar Hall effects, in Cartesian (left) and polar (right) coordinates. For the conventional PHE, both the longitudinal (anisotropic magnetoresistance, $R_{xx}$, blue) and transverse (planar Hall effect, $R_{yx}$, red) resistances exhibit a $\pi$-periodic angular dependence, with a $\pi/4$-offset between them. The origin of the oscillation is set by the direction of the electric field (current). For the APHE, the angular dependence is $2\pi/3$-periodic, with origin set by the crystal directions, and is not associated with any AMR. 
			\textbf{c}: Crystal structure of trigonal-\ce{PtBi2}, with layered nature and in-plane $\mathcal{C}_{3}$-symmetry highlighted. 
			\textbf{d}: Sample configuration. The angle $\varphi$ refers to the orientation of the in-plane magnetic field \textbf{B}.}
		\label{Fig1_Principle}
	\end{figure*}
	
	In low-dimensional systems, such as \ce{LaAlO3}/\ce{SrTiO3} oxide interfaces, the occurrence of an APHE has been reported and is due to a Zeeman-induced modification of local concentrations of the out-of-plane Berry curvature, which, when integrated over momenta, becomes non-vanishing~\cite{Lesne2023}. An APHE has also been reported in \ce{VS2}-\ce{VS} heterostructures~\cite{Zhou2022}. The mechanism we find to be at work in TNLs is completely different in nature.
	It is caused by a Zeeman-induced conversion of TNLs into Weyl nodes that generalizes the fusion of Weyl nodes into nodal lines predicted to occur in mirror-symmetric systems~\cite{Sun2018}. The important point is that  magnetic fields that break the mirror symmetry protecting the TNLs lead to a nonlocal conversion of the TNL into Weyl nodes of opposite chirality, meaning that they are separated in momentum space by a vector that has components parallel to the mirror plane. The extraordinary feature of this $k$-space separation in momentum space is that it survives even for infinitesimally small magnetic field and can be as large as the diameter of the TNL. This conversion and its properties can be qualitatively captured using a simple two-band low-energy model (see SM).
	This generally induces large momentum regions of non-zero Chern number, thus generating an anomalous planar Hall effect (APHE) already at infinitesimal magnetic field, with a much larger amplitude than that which would be obtained solely from the Zeeman-induced displacement of Weyl nodes (see SM).
	%
	Consider for simplicity a single pair of TNLs related to each other by time-reversal symmetry and protected by a vertical mirror plane, which, without loss of generality, we set as ${\mathcal M}_x$ (see Fig. \ref{Fig1_Principle}.a). With an infinitesimally small magnetic field along the $\hat{y}$ direction, the TNLs converts into two field-induced pairs of Weyl nodes, each of which has a separation in $k_z$ comparable to the TNL dimension, generating a dissipationless (i.e. without diagonal components) antisymmetric Hall conductance $\sigma_{yx}$ (i.e. an APHE). The system can be viewed in fact as a collection of two dimensional $\left\{k_x,k_y \right\}$ insulating layers~\cite{lau17} parameterized by $k_z$ and characterized by a local Chern number $c(k_z)$ (see Fig. \ref{Fig1_Principle}.a) that is changed by the topological charge of each Weyl node. The nonlocal Zeeman-induced conversion of the TNLs into Weyl nodes then leads to a net $\sigma_{yx}=\int c(k_z) d k_z$ and to Hall voltages that lie in the same plane as the applied current and the external magnetic field, precisely in a configuration where the conventional Hall effect is absent. Such APHE thus represents a characteristic electromagnetic response of TNLs. 
	
	Additionally, Onsager's relations~\cite{ons31} enforce the transversal APHE conductance to be odd under a magnetic field reversal and thus compatible only with an out-of-plane threefold rotational symmetry (see Fig. \ref{Fig1_Principle}.b).
	This property, along with its non-dissipative character, makes the APHE experimentally distinguishable from a planar Hall effect. First, a PHE, such as that associated with the Berry curvature of the Weyl nodes, is characterized by a $\pi$-periodic oscillation of both the longitudinal $R_{xx}$ and transverse resistance $R_{yx}$, when the magnetic field is rotated in-plane while keeping the current direction fixed, with a $\pi/4$ offset between them (see Fig. \ref{Fig1_Principle}.b and Methods) ~\cite{Burkov2017,Nandy2017}. The longitudinal resistance $R_{xx}$ is moreover maximized when magnetic field and current are aligned, while the transverse resistance $R_{yx}$ vanishes in this configuration. As a result, a two-fold symmetric PHE aligned with current direction can easily be disentangle from a threefold-symmetric APHE aligned with crystaline axes (see Fig. \ref{Fig1_Principle}.b). More importantly, the APHE is non-dissipative, i.e it is not associated with any corresponding longitudinal signal. This allows to distinguish it unambiguously from any potential three-fold symmetric PHE due to magneto-crystalline anisotropies, which would be associated with a corresponding AMR.
	
	\section*{Anomalous planar Hall effect}
	
	\begin{figure*}[t]
		\centering
		\includegraphics[width=1\textwidth]{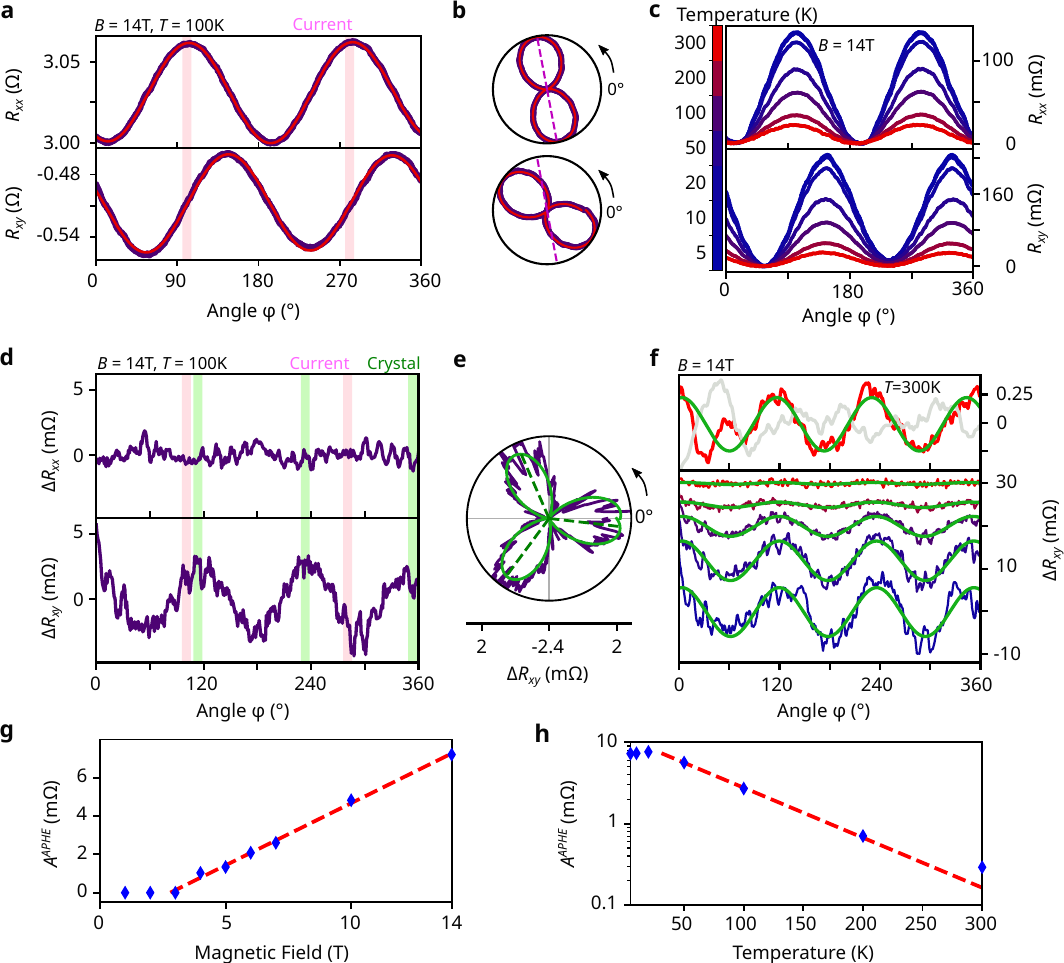}
		\caption{\textbf{Standard and Anomalous Planar Hall effect in \ce{PtBi2}}. \textbf{a,b}: Angular dependence of $R_{xx}$ and $R_{yx}$ at 14T, 100K in cartesian (\textbf{a}) and polar coordinates (\textbf{b}). The fits with Equation 1 (Methods) are shown in red in \textbf{a} and \textbf{b}. The radial axis has the same range as in \textbf{a}.  The pink bars in \textbf{a} and pink dashed line in \textbf{b} show the current direction estimated from the fits, with a $\pm 5\degree$ width. 
			\textbf{c}: Angular dependence of $R_{xx}$ and $R_{yx}$ at different temperatures from 5K to 300K, at 14T. The curves are vertically shifted for clarity.
			\textbf{d,e}: Angular dependence of the residues $\Delta R_{xx}$ and $\Delta R_{yx}$ from the data in \textbf{a} after a background removal (see Supplementary sec. \ref{SI sec: simplefit_B}), in cartesian (\textbf{d}) and polar coordinates (\textbf{e}). A $2\pi/3$-periodic signal is clearly visible in $\Delta R_{yx}$. The pink and green bars show the previously estimated current direction and the crystal direction estimated from the fits to Supplementary Materials eq. S5, respectively, with a $\pm 5\degree$ width. In \textbf{e}, the fit to Supplementary Materials eq. S5 is shown in green. 
			\textbf{f}: Bottom: angular dependence of $\Delta R_{yx}$ at 14T for $T = $\, 20K, 50K, 100K, 200K and 300K, with fits to Supplementary Materials eq. S5 shown in green. The curves are vertically shifted for clarity. Top: Angular dependence of $\Delta R_{yx}$ at 14T, 300K, with fit in green. The 300K data was smoothed over 31° for visibility. The corresponding $\Delta R_{xx}$ signal is plotted in grey line with the same scale for comparison, and shows no visible periodic signal. 
			\textbf{g}: Field dependence of the APHE amplitude $A^{\text{APHE}}$, showing a linear dependence above $B_{c} \sim 2.8$~T. \textbf{h}: Temperature dependence of $A^{\text{APHE}}$, showing an exponential decay above $T_c \sim 30$~K, with an energy scale $\Delta \sim 6$~meV.}
		\label{Fig2_Experiment}
	\end{figure*}
	
	We now show that both these effects can be probed in the layered van der Waals material \ce{PtBi2}, that 
	has recently been characterized as a non-magnetic type I Weyl semimetal. \ce{PtBi2} also exhibits sub-Kelvin 2D-superconductivity and a BKT transition in nanostructures~\cite{Veyrat2023}, as well as higher-temperature surface superconductivity~\cite{Schimmel2023} with superconducting topological Fermi arcs~\cite{Kuibarov2024}. The crystallographic point group symmetry of \ce{PtBi2} is ${\mathcal C}_{3v}$ that is comprised of a three-fold axis and three vertical mirror planes ${\mathcal M}$ (
	Fig. \ref{Fig1_Principle}.c)~\cite{Shipunov2020}, and is compatible with the appearance of an APHE. When an in-plane magnetic field is perpendicular to a mirror plane, the anomalous Hall conductivity (AHC) $\sigma_{yx}$ must vanish. Conversely, when the field is parallel to a mirror plane, $\sigma_{yx}$ is maximal~\cite{Battilomo2021}. This results in a $2 \pi /3$ periodic angular dependence (see Fig. \ref{Fig1_Principle}.c) that can be detected in practice with magnetotransport measurements. 
	
	We focus our study on a 70nm thick nanostructure (see magnetotransport measurement schematic in Fig. \ref{Fig1_Principle}.d) investigated up to 14T, and temperatures from 5K up to 300K (another structure showed similar behavior, see Supplementary Material). The results are shown in Fig. \ref{Fig2_Experiment}.
	First, in exfoliated nanostructures of \ce{PtBi2}, we systematically observed a PHE. At $T=100$~K and $B=14$~T, a pronounced $\pi$-periodic oscillation is clearly visible in both $R_{xx}$ and $R_{yx}$ (Fig. \ref{Fig2_Experiment}.a), with the expected $\pi/4$ angular shift between them (Fig. \ref{Fig2_Experiment}.b). The PHE is already visible at magnetic fields as low as 1 T (see Supplementary Materials Fig. \ref{SI: FigSI3}). 
	The angular positions of the maxima of $R_{xx}$ are consistent with the expected current orientation in the sample (see Methods). The PHE is very robust with temperature, and for $B=14$~T it can be evidenced up to room temperature (Fig. \ref{Fig2_Experiment}.c).
	The presence of a strong PHE in the non-magnetic \ce{PtBi2} reveals the large BC present in the material, giving significant experimental indications of its Weyl nature, and confirming predictions from previous band structure calculations~\cite{Veyrat2023}. 
	
	The main experimental result of this work is the evidence of an APHE in our measurements, which appears as a small deviation from the PHE in Fig. \ref{Fig2_Experiment}.a-c. In order to evidence the APHE in our data, we remove a $\pi$- and $2\pi$-periodic background from each measurement (in red in Fig. \ref{Fig2_Experiment}.a,b, and Supplementary sec. \ref{SI sec: simplefit_B}). Removing each signal separately would be similar to antisymmetrizing (resp. symmetrizing) the data in magnetic field, albeit with more control over exactly which terms are removed. The resulting residues are depicted in Fig. \ref{Fig2_Experiment}.d-f. At 14T, $2\pi/3$-periodic oscillations, which contrary to the standard PHE signal are antisymmetric in B, are clearly visible in the transverse resistance residues $\Delta R_{yx}$ at 100K (Fig. \ref{Fig2_Experiment}.d,e) and can be fitted with cosine fits (Fig. \ref{Fig2_Experiment}.e, in green) . This $2\pi/3$-periodic signal appears above $B = 4$T, at a constant angular position, and its amplitude increases with magnetic field (Supplementary sec. \ref{SI sec: APHE vs B}). 
	Importantly, no associated $2\pi/3$-periodic signal is visible in the longitudinal resistance residues $\Delta R_{xx}$, up to the highest fields and down to the lowest temperatures (Supplementary Fig. \ref{SI: FigSI5}). This lack of longitudinal component is critical in distinguishing the dissipationless APHE from any conventional, dissipative PHE which would necessarily be associated with an AMR. Remarkably, the APHE is very robust in temperature, as the oscillations remain visible from 5K up to room temperature, as shown in Fig. \ref{Fig2_Experiment}.f.
	
	The field and temperature dependence of the APHE signal obtained from the cosine fits are presented in Fig. \ref{Fig2_Experiment}.g,h. At 5K, the APHE signal is visible above 4T, and increases linearly with field. This fit yields a critical magnetic field $B_c \sim 2.8$~T for the appearance of an APHE.  At 14T, the APHE signal remains constant in temperature below 20K, similar to the PHE and the longitudinal resistance (see Supplementary sec. \ref{SI sec: simplefit_T}). Above this temperature, it decreases following an exponential decay law $A^{APHE}(T) - A^{APHE}(T=0) \propto e^{-k_B(T-T_c)/\Delta}$ with $T_c \sim 30$~K an an energy scale $\Delta \sim 6$~meV. 
	The existence of an onset field for the APHE could be explained by the fact that, while the TNLs are already topologically gapped at infinitesimal fields, the APHE would only become visible when the gap of the TNL exceeds the thermal activation energy.
	Similarly, the exponential decay of the signal in temperature could come from thermal activation through the TNL gap opened by the magnetic field. This is supported by the temperature shift on the onset field evidenced in sample D3 (see Fig. \ref{SI: FigSI8b}).
	In this case, the energy scale of the decay, $\Delta$, would be linked to the gap of the TNLs.
	
	
	\section*{Topological nodal lines in \ce{PtBi2} from DFT}
	
	\begin{figure*}[t!]
		\centering
		\includegraphics[width=0.4\textwidth]{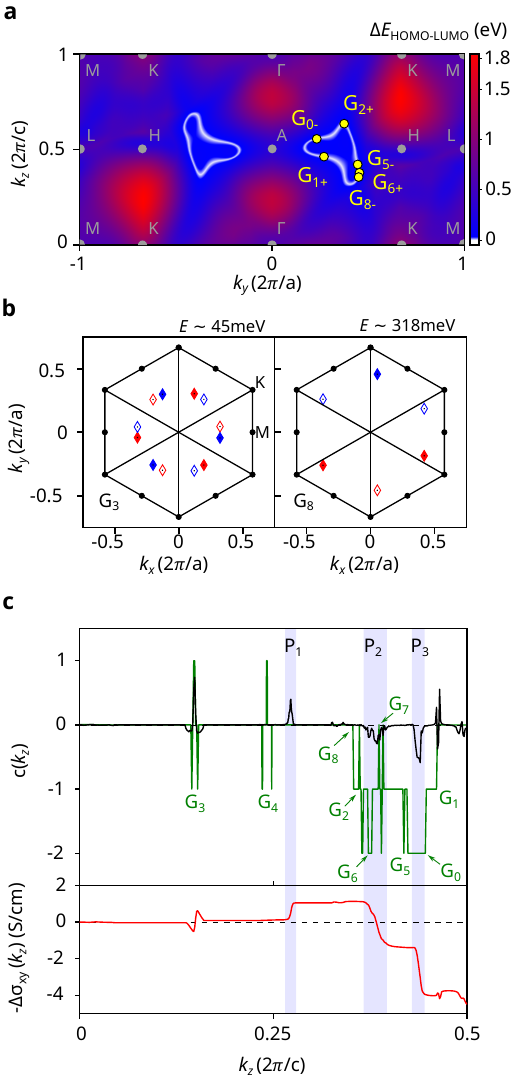}
		\caption{\textbf{Nodal lines and the origin of the anomalous Planar Hall effect in \ce{PtBi2}}. 
			\textbf{a}: Energy gap $\Delta E$ between HOMO and LUMO bands in the $k_y, k_z$ (mirror) plane. The nodal loops ($\Delta E = 0$) appear in white. When $B \neq 0$, each nodal loop splits into 6 Weyl nodes (WN, yellow points), forming 6 groups of 6-WN. The signs denote the chiralities. 
			\textbf{b}: Two groups of WN of HOMO-LUMO for a Zeeman energy $E_Z = 14$~meV: $G_3$ is the 12-fold set of WNs closest to $E_F$ already present at $B=0$, and $G_8$ is one of the six 6-fold groups mentioned above. The average energies of the groups are shown. Red (blue) markers denotes positive (negative) chirality, while full (empty) markers denote the positive (negative) $k_z$ position of the WN ($G_3: k_z \sim \pm 0.149$, $G_8: k_z \sim \pm 0.358$). Solid lines represent the mirror planes, while the dots show the high-symmetry points.
			\textbf{c}: (Top) Chern number $c(k_z)$ in an ideal (green, full HOMO) and a more realistic (black, $E_F = E_{G_3} = 45.3$~meV) case, with a a Zeeman energy $E_Z = 14$~meV. In the ideal case, the Chern number jumps discretely by $\pm 1$ at each WN, while the variation is smoothed out in the realistic case. (Bottom) Anomalous Hall conductivity $-\Delta \sigma_{xy}(k_z)$ calculated from the Chern signal in the realistic case (in black above). The 12 WNs from $G_3$ at low $k_z$ contribute very little to the AHC, as the Berry curvature they generate is nearly compensated. Most of the AHC comes from 2 peaks in the Chern number at higher $k_z$, $P_2$ and $P_3$ (shown in blue). A third peak at lower $k_z$, $P_1$, attenuates the total AHC amplitude, and is found to correspond to WNs from nodal lines below the HOMO band (see Supplementary materials sec. \ref{SI subsec: band47}). Only the $k_z >0$ dependences are shown, as $c(k_{z})$ is even and $\Delta \sigma_{xy}$ is odd in $k_z$.}
		\label{Figure3_Theory}
	\end{figure*}
	
	We next demonstrate that the APHE experimentally observed originates from the presence of TNLs in \ce{PtBi2} by performing full-relativistic electronic band structure calculations, with and without magnetic field. 
	In the absence of Zeeman coupling the material features 3 groups of Weyl nodes, which, due to the concomitant presence of time-reversal symmetry and the threefold rotation symmetry, all come with multiplicity twelve. These groups of Weyl nodes appear above the Fermi energy, with the one lowest in energy being at around 45.3~meV above $E_F$ , in agreement with a previous study~\cite{Veyrat2023}. 
	In each of these groups pairs of Weyl nodes of opposite chirality are connected by a vector perpendicular to the mirror plane. The presence of in-plane magnetic field leads to a movement of the Weyl nodes in all momentum directions due to the low residual symmetry: the system possesses at most a vertical ${\mathcal M}^{\prime}= {\mathcal M} \times \Theta$ symmetry (with $\Theta$ time-reversal)  when the magnetic field is parallel to the mirror plane ${\mathcal M}$. 
	However, the Weyl node displacement is proportional to the strength of the applied magnetic field, and remains relatively small at laboratory-accessible fields. The resulting anomalous planar Hall conductance from this displacement is therefore expected to be vanishingly small. 
	
	The situation is completely different for the three pairs of nodal loops revealed by our calculations, which lie in the vertical mirror planes of \ce{PtBi2} and are reminiscent of nodal chain semimetals~\cite{Sun2018}. An infinitesimal magnetic field leads to the conversion of the TNLs, each into 6 Weyl nodes [see Fig. \ref{Figure3_Theory}.a]. Since in \ce{PtBi2} the degeneracy loops do not occur at fixed energies, these Zeeman-induced Weyl nodes form different groups that are separated in energy -- we remark that one of these groups has a complex evolution as the magnetic field is increased as it directly combines with one of the preexisting twelvefold Weyl node groups at $B=0$ (see Supplemental Material). For a magnetic field parallel to one mirror plane  each of these groups is sixfold, with pairs of opposite chirality related by the combined ${\mathcal M}^{\prime}$ symmetry. This immediately implies that an isolated group yields a sizable contribution to the anomalous planar Hall conductance. The latter can be computed in an "ideal" case by simply assuming that the Weyl nodes are all at the Fermi level, in which case each Weyl node provides a unit change to the Chern number of the insulating $\left(k_x,k_y \right)$ layers. 
	The distribution of the Weyl nodes in two trios (see Fig. \ref{Figure3_Theory}.b) at nearly opposite values of $k_z$ demonstrates that a large contribution to the anomalous planar Hall conductance can be expected from the sixfold groups of Zeeman-induced Weyl nodes. This is verified by a direct calculation of the local Chern number of the full system assuming, as before, all Weyl nodes are at the Fermi energy and a Zeeman energy $E_Z = 14$~meV from an applied magnetic field parallel to one of the vertical mirror planes of the material (see Fig. \ref{Figure3_Theory}.c, in green). We note that $E_Z$ was chosen at such a value for numerical resolutions purposes, and does not correspond to our experimental conditions.
	As expected, the twelvefold groups of precursive Weyl nodes do not give a large contribution to the anomalous planar Hall conductance.
	
	We have also performed a realistic calculations assuming that bands are filled with a Fermi level set at the energy of the Weyl nodes group at ~45.3~meV  above the non-magnetic $E_F$ (black line in Fig. \ref{Figure3_Theory}.c). We find that the local Chern signal (see Supplemental Material) of the twelvefold groups is washed out whereas the jumps due to the sixfold groups originating from the conversion of the TNLs are smoothed. 
	However, the realistic contribution to the anomalous planar Hall conductance (red line in Fig. \ref{Figure3_Theory}.c) is not only due to these TNLs but also derives from the existence of additional nodal lines (see Supplemental Material) that we find in lower valence bands and also undergo conversion into Weyl nodes in the presence of planar magnetic fields. This further demonstrates that the anomalous planar Hall conductance of \ce{PtBi2} is a direct electromagnetic response of TNLs.

	
	\section*{Discussion and outlooks}
	To sum up, we have demonstrated that 3D nanostructures of the non-magnetic 3D Weyl semimetal \ce{PtBi2} exhibit, beyond a conventional PHE, a very robust APHE with a signature $2\pi/3$-periodic oscillation in $\Delta R_{yx}$ and an absent dissipative signal in $\Delta R_{xx}$. 
	We have shown that this APHE originates from previously unidentified topological nodal lines in \ce{PtBi2}'s band structure, through a non-local conversion to Weyl nodes under a magnetic field.
	This mechanism can be generally used to engineer Weyl nodes in materials featuring mirror symmetry-protected nodal lines by means of arbitrary small magnetic fields. Our observations also establish
	%
	the anomalous planar Hall effect as an efficient magneto-transport tool to reveal the presence of TNLs in trigonal semimetals, which could so far only be characterized through spectroscopy measurements. 
	%
	Demonstrating the presence of topological features through transport is especially interesting in \ce{PtBi2}, where 2D superconductivity was recently reported~\cite{Veyrat2023}, and a recent ARPES study further found the superconducting weight to be localised on the topological Fermi arcs~\cite{Kuibarov2024}, opening perspectives for possible topological superconductivity.

	
	\bibliography{Veyrat_2024_APHE}
	
	\bibliographystyle{apsrev4-2}


	\section*{Acknowledgments}\label{sec: Acknowledgments}
	A.V. acknowledges funding from the European Research Council (ERC) under the European Union’s Horizon 2020 research and innovation program (grant Ballistop agreement no. 833350). S.A. acknowledges the financial support of (DFG) through the grant AS 523/4-1. C.O. acknowledges support from the MAECI project “ULTRAQMAT”. L.V. was supported by the Leibniz Association through the Leibniz Competition.  This work was supported by the Deutsche Forschungsgemeinschaft (DFG, German Research Foundation) through the Sonderforschungsbereich SFB 1143 and under Germany's Excellence Strategy through the W\"{u}rzburg-Dresden Cluster of Excellence on Complexity and Topology in Quantum Matter -- \emph{ct.qmat} (EXC 2147, project-ids 390858490 and 392019.

	\clearpage
	
	\section*{Supplementary materials}
	
	\subsection*{Methods}\label{sec: Methods}
	\subsubsection*{Sample Preparation}
	High-quality single crystals of \ce{PtBi2} were grown using the self-flux method \cite{Shipunov2020}. These crystals were mechanically exfoliated to obtain thin flakes, with widths exceeding $10~\mu$m and thickness ranging from a few dozen to a few hundred nanometers. The flakes were contacted with Cr/Au using standard e-beam lithography techniques. Prior to the metal deposition, a small Ar-etch was performed to eliminate any surface oxidation.
	The main sample used in this study is denoted as $D1$ (70 nm thick), and supplementary information includes corroborating results for a second sample, D2 (126 nm thick). 
	In a previous study\cite{Veyrat2023}, the two dimensional superconductivity of these samples were studied in details at sub-Kelvin temperatures. Here, we focus on measurements performed above 1K, above the superconducting transition. No evidence of aging effects was observed between the two studies, as indicated by the absence of any measurable change in the residual resistance ratio ($RRR = R(300K)/R(4K)$).
	
	\subsubsection*{Measurement Setup}
	The measurement configuration consists of a standard Hall-bar geometry. A current is injected between the source and the drain as depicted in Fig. \ref{Fig1_Principle}.d. 
	Longitudinal and transverse resistances (indicated in red and black, respectively) are measured along and across the sample relative to the current orientation. 
	
	\textbf{Measurement set-up for samples D1 and D2.}
	
	Measurements were conducted in a Dynacool 14T PPMS using an insert equipped with a mechanical 2D rotator.
	By rotating the sample with the rotator, the angle $\varphi$ between the fixed-axis magnetic field and the applied current can be adjusted over a full range of 360° (with $\varphi$ the angle between the magnetic field and the electric field. 
	The resistances were measured using external lock-in amplifiers, with an AC current of 100~$\mu$A at a frequency of 927.7~Hz, with an integration time of~300 ms. At such low currents, no thermal effects are expected. For sample $D1$, for measurements taken at $T=5$~K and $B=1,2,3,4,5,6,7,10$~T, as well as at $B=14$~T and $T=5,10,20,50,300$~K, 10 points were measured at each angular position, taking the averaged value of the resistance. The angular step for each measurement was 1°. All measurements on sample $D2$ were conducted with the same parameters. 
	For sample $D1$, more precise measurements were taken at $T=5$~K and $B=14$~T, as well as at $B=14$~T and $T=100,200$~K, with an averaging over 40 measurement points at each angular position. The angular step for each measurement was 0.5°, and the results were interpolated with a step of 1°, to perform the analysis in the same way for each pair of (B,T) parameters.

	At low temperature ($T \leq 20$K) the first oscillation of the APHE ($ 0\degree \leq \varphi \leq 120\degree$) is not fully visible even at 14T, although it is consistently observed for $T \geq 50$~K (see at $T=100$K in Fig. \ref{Fig2_Experiment}.a and Supplementary sec. \ref{SI sec: APHE vs T} and sec. \ref{SI sec: APHE vs B}). 
	This partial suppression of the signal likely stems from the mechanical rotator: When the stepper-motor at the top of the measurement stick turns by a small angle (in our measurements, the angular step is 1°), the mechanical rotator in the cryostat will move by an inconsistent angle (around the target step, e.g. 1°). 
	As we measure the angle of the rotator at the top of the stick, and not the actual angle of the sample at the bottom, this creates small deviations of the PHE signal away from a $\pi$-periodic oscillation. When a $\pi$-periodic background is removed from the data (to evidence the APHE signal), these deviations are carried to the residues, and can corrupt the signal. 
	These artifacts are reproducible, and decrease with temperature, as would be expected with mechanical rotator inconsistencies.
	
	\textbf{Measurement set-up for sample D3.}
	
	Measurements were conducted in a VTI equipped with a 3D-piezorotator and a large bore 14T magnet. In this work, we show in-plane rotation measurements.
	By rotating the sample with the rotator, the angle $\varphi$ between the fixed-axis magnetic field and the applied current can be adjusted over a 180° range (with $\varphi$ the angle between the magnetic field and the electric field). In order to have the full 360° range, we flip the orientation of the sample by 180° along the perpendicular axis. This is equivalent to reverse the direction of the magnetic field.
	The resistances were measured using external lock-in amplifiers, with an AC current of 500~$\mu$A at a frequency of 331~Hz, with an integration time of~300 ms. For sample $D3$, for measurements taken at $T=3, 20, 100$~K for fields ranging between 1~T and 14~T. Contrary to the measurement of D1 and D2, a single point was measured at each angular position. 

	\subsubsection*{Planar Hall Effect}
	\label{sec: PHE}
	
	The contributions of the planar Hall effect/anomalous magnetoresistance to the longitudinal resistivity $\rho_{xx}$ and transverse resistivity $\rho_{yx}$ obey the following angular dependence~\cite{Burkov2017,Nandy2017}:
	\begin{align} \label{eq: PHE}
		\begin{split}
			\rho^{\text{AMR}}_{xx}(\varphi) &= \rho_\perp - \Delta\rho \: \cos^2{\varphi} ,\\ 
			\rho^{\text{PHE}}_{yx}(\varphi) &= - \Delta\rho \cos\varphi \sin \varphi,    
		\end{split}
	\end{align}
	with $\Delta \rho = \rho_\parallel - \rho_\perp$ the amplitude of both the PHE and the AMR ; $\rho_\parallel$ and $\rho_\perp$ the resistivities when \textbf{B} is respectively along and perpendicular to the electrical field (current); and $\varphi$ the angle between the magnetic and electric fields (i.e. current lines) in the sample.
	The PHE signal is therefore characterized by $\pi$-periodic oscillations for both $\rho_{xx}$ and $\rho_{yx}$ (when rotating the magnetic field in the sample's plane, with a fixed current) with the same amplitude, with a $\pi/4$ offset between the two. The maxima of $\rho_{xx}$ correspond experimentally to the orientation of the current in the sample.
	
	\subsubsection*{Computation details}\label{sec:computationaldetails}
	We performed a full-relativistic non-magnetic calculation using the
	full potential local orbital (FPLO) code \cite{Koepernik1999} version 22.01
	within the generalized gradient approximation (GGA) \cite{Perdew1996}.
	From the DFT result a 72-band Wannier function (WF) model was extracted
	consisting of Bi$6p$ and Pt$6s5d$ type WFs. A constant magnetic
	field Zeeman term $H^{\mathrm{Zeeman}}=\boldsymbol{B}\mu_{B}\left\langle \boldsymbol{S}\right\rangle $
	was added to the model using the WF representation of the spin operators
	$\left\langle \boldsymbol{S}\right\rangle $.

	The anomalous planar Hall signal $\Delta\sigma_{yx}$ can be computed
	by calculating the Chern signal $c\left(k_{z}\right)=\frac{1}{2\pi}\int F_{z}
	\left(\boldsymbol{k}\right)d\boldsymbol{S}$
	for a number of $k_{z}$-planes with subsequent integration over $k_{z}$.
	For an assumed constant homo $c\left(k_{z}\right)$ can
	be obtained by a plaquette type integration (see sec. \ref{SI subsec:Chern-signal} and Ref.~\cite{Fukui2005})
	and with a Fermi level by a simple Riemann-sum integral.

	\onecolumngrid
	\newpage

	\subsection*{Materials}
	
	\renewcommand{\figurename}{Figure}
	\renewcommand{\thefigure}{S\arabic{figure}}
	\renewcommand{\theequation}{S\arabic{equation}}
	\renewcommand{\thesubsubsection}{S\arabic{subsubsection}}
	\setcounter{figure}{0}
	
	\subsubsection{Resistivity in the anomalous planar Hall effect} \label{SI sec: rho APHE}
	
\begin{figure*}[ht]
		\centering
		\includegraphics[width=\textwidth]{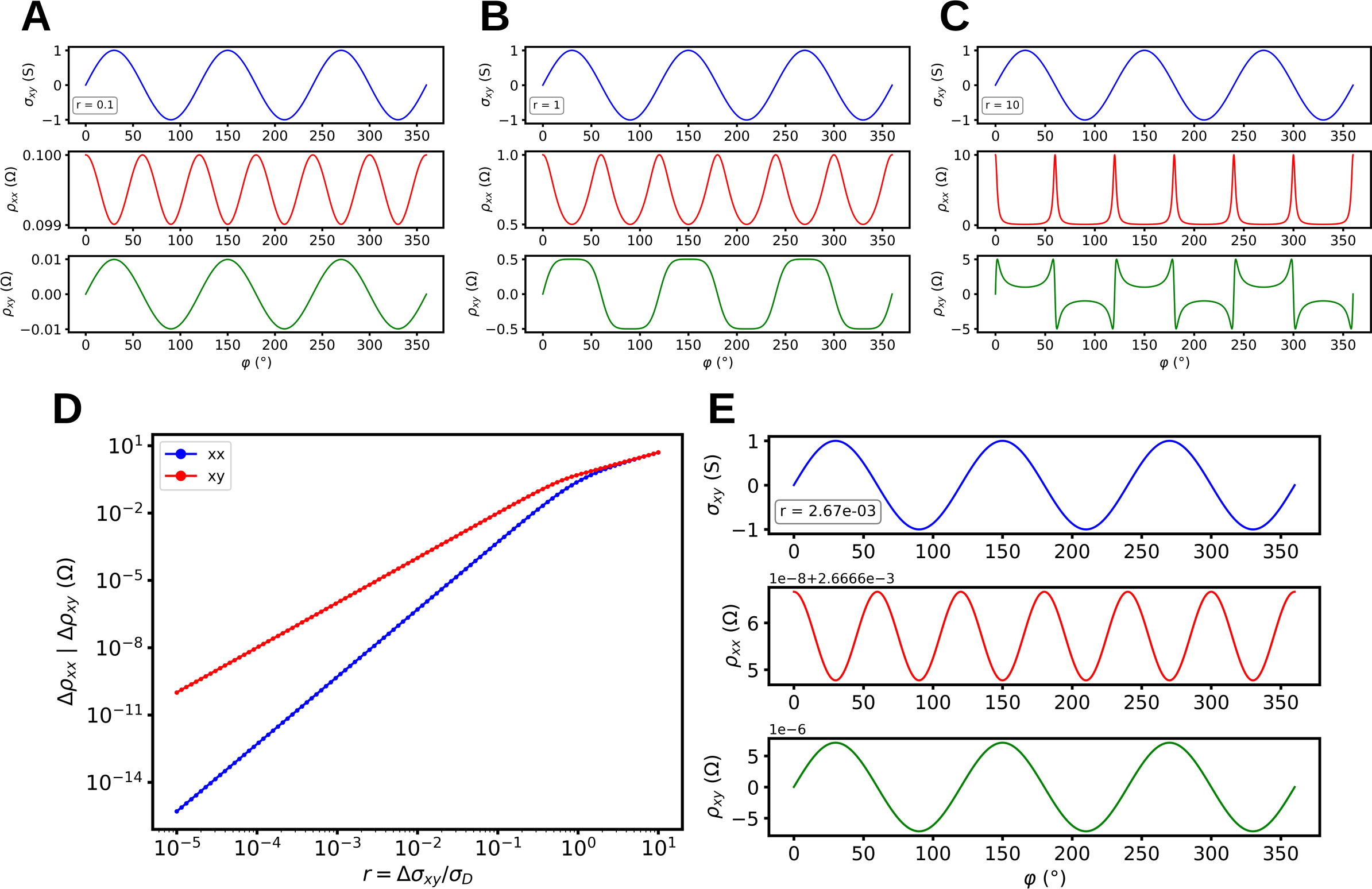}
		\caption{Resistivitities of the APHE. a,b,c: $\rho_{xx}$ and $\rho_{xy}$ obtained from $\sigma$ for the ratios 0.1,1 and 10. d: Amplitude of the oscillations of $\rho_{xx}$ and $\rho_{xy}$ with respect to the ratio $r$, in log scale. e: $\rho_{xx}$ and $\rho_{xy}$ obtained for a realistic value of $r = 2.67 \times 10^{-3}$.}
		\label{SI: FigSI rho APHE}
		
	\end{figure*}

	It is expected from the theory~\cite{Battilomo2021} that the conductivity generated by the anomalous planar Hall effect (APHE) follows the angular dependence
	\begin{align} \label{SI eq: APHE AHC}
		\begin{split}
			\sigma^{\text{APHE}}_{xx}(\varphi) &= 0,\\ 
			\sigma^{\text{APHE}}_{yx}(\varphi) &= \Delta \sigma_{yx} \cos 3\varphi.    
		\end{split}
	\end{align}
	In order to obtain the angular dependence of the resistivity associated with the APHE, we assume an underlying constant (Drude) longitudinal conductivity $\sigma_{xx} = \sigma_D$, so that the overall conductivity matrix becomes
	$\sigma = \begin{pmatrix}
		\sigma_D & \sigma_{xy}\\
		-\sigma_{xy} & \sigma_D
	\end{pmatrix}$, with $\sigma_{xy} = \sigma^{\text{APHE}}_{xy}(\varphi)$ from eq. S1. By inverting the conductivity matrix, we obtain the resistivity matrix $  \rho = \sigma^{-1} =\begin{pmatrix} \rho_{xx} & \rho_{xy}\\ -\rho_{xy} & \rho_{xx}  \end{pmatrix}$, with
	\begin{align} \label{SI eq: APHE rho_sigma}
		\begin{split}
			\rho_{xx} &= \dfrac{\sigma_D}{\sigma_D^2 + \sigma_{xy}^2} = \dfrac{1}{\sigma_D(1 + (\sigma_{xy}/\sigma_D)^2)},\\ 
			\rho_{xy} &= -\dfrac{\sigma_{xy}}{\sigma_D^2 + \sigma_{xy}^2} = \dfrac{\sigma_{xy}/\sigma_D}{\sigma_D(1 + (\sigma_{xy}/\sigma_D)^2)}.    
		\end{split}
	\end{align}
	Here, we can see that the APHE has a more complicated effect on the resistivity than on the conductivity, with both $\rho_{xx}$ and $\rho_{xy}$ having $2\pi/3$-periodic components in $\varphi$. The shape of the oscillation depends strongly on the relative amplitude $r =\Delta \sigma_{xy} / \sigma_D$ of the AHC oscillation with respect to the longitudinal conductivity (see Fig. \ref{SI: FigSI rho APHE}.A,B,C). For $r \ll 1$, both $\rho_{xx}$ and $\rho_{xy}$ tend towards sine waves of period $\pi/3$ and $2\pi/3$, respectively. The amplitudes of the oscillations of $\rho_{xx}$ and $\rho_{xy}$ (defined as $\Delta \rho = [\text{max}(\rho)-\text{min}(\rho)]/2$) depend on the ratio $r$ as $r^3$ and $r^2$, respectively, meaning that for $r \ll 1$, the oscillations in $\rho_{xx}$ will be much lower than those in $\rho_{xy}$ (see Fig. \ref{SI: FigSI rho APHE}.F). 
	
	As the ratio $r$ is proportional to the angular mean value $\overline{\rho}_{xx}$ of $\rho_{xx}$, this means that for a given amplitude of the oscilations in AHC, the APHE in resistivities will deviate from sine waves when the longitudinal resistivity is too high, and the oscillations will vanish if the longitudinal resistivity is too low. For our sample, we can roughly estimate a ratio $r = \dfrac{\Delta \sigma_{xy}}{\sigma_D} \sim 2.7 \times 10^{-3}$, by taking $\sigma_D = 1/\overline{\rho}_{xx} = 1/ (R_{xx}(B=0)/N_\square) \sim  1/ (0.8/0.3)$ and $\Delta \sigma_{xy} = 0.001$, which gives an amplitude of the oscillations of $\rho_{xy}$ of about 7$\mu\Omega$, close to what we see at 14T (see Fig. \ref{Fig2_Experiment}.d). At this value, the oscillations in $\rho_{xy}$ are very close to a sine wave, and the oscillations in $\rho_{xx}$ are about 3 orders of magnitude smaller. This shows that, in our case, we can simply treat the signature of the APHE in resistivity as
	\begin{align} \label{SI eq: APHE rho}
		\begin{split}
			\rho^{\text{APHE}}_{xx}(\varphi) &= 0,\\ 
			\rho^{\text{APHE}}_{xy}(\varphi) &= A^{\text{APHE}} \cos 3\varphi.    
		\end{split}
	\end{align}

	\clearpage \subsubsection{Residues extraction for APHE vs B} \label{SI sec: simplefit_B}
	\begin{figure*}[ht]
		\centering
		\includegraphics[width=\textwidth]{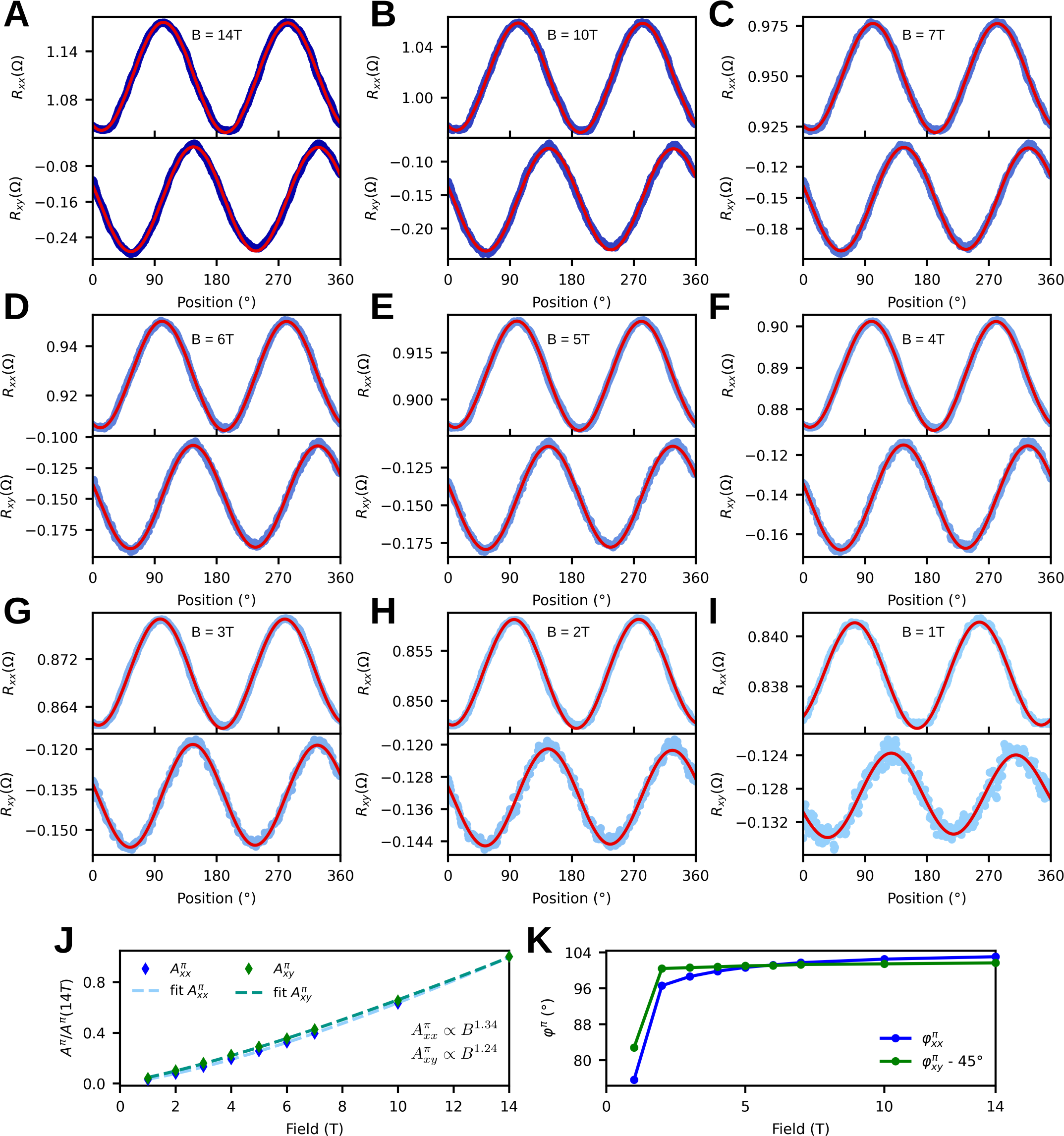}
		\caption{}
		\label{SI: FigSI3}
		\title{Raw data and simple fits at 5K, between 1T and 14T}
	\end{figure*}
	
	In order to extract as much of the background and PHE as possible, and try to keep only the APHE, we can fit $R_{xx}$ and $R_{yx}$ separately at each field with a simple model:
	
	\begin{align} \label{SI eq: simple_fit}
		\overline{R}(\varphi) =  C + A^{2\pi} \cdot \text{cos} \: ( \varphi - \varphi_{2\pi}) + A^{\pi} \cdot \text{cos} \: [2 (\varphi - \varphi_{\pi})],   
	\end{align}
	with $C$ an angle-independent constant, and $A^{\pi}$ and $A^{2\pi}$ the amplitudes of some $\pi$- and $2\pi$-periodic signals, with angular origins $\varphi_{\pi}$ and $\varphi_{2\pi}$, respectively. The $\pi$-periodic component should remove the PHE signal from the measurements, while the angle-independent and $2\pi$-periodic components should remove the magnetoresistance signal from the perpendicular field component (respectively for the misalignment between the rotation plane and magnetic field axis, and for the misalignment of the sample's plane with the rotation plane). 
	All the above parameters are independent between $R_{xx}$ and $R_{yx}$, and are assumed to be field- or temperature-dependent. Fits of the data at each magnetic field are shown in red in Fig. \ref{SI: FigSI3}.A-I. The residues $\Delta R$ are then obtained by subtracting the previous fit from the data: $\Delta R = R - \overline{R}$. \\

	\clearpage \subsubsection{Residues extraction for APHE vs T} \label{SI sec: simplefit_T}
	\begin{figure*}[ht]
		\centering
		\includegraphics[width=\textwidth]{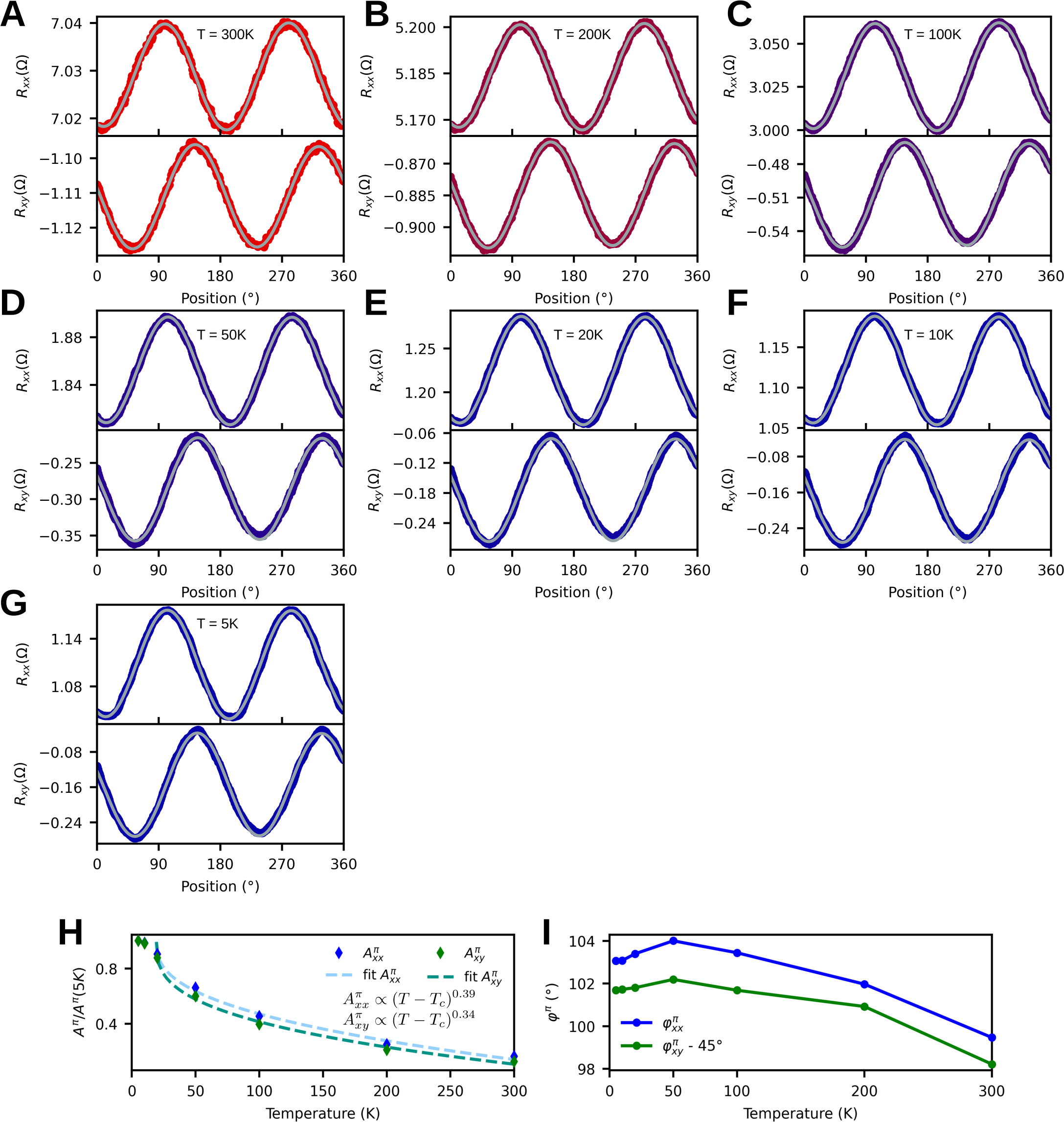}
		\caption{}
		\label{SI: FigSI4}
		\title{Raw data and simple fits at 14T, between 5K and 300K}
	\end{figure*}
	
	We perform the exact same analysis as in sec. \ref{SI sec: simplefit_B} at 14T and multiple temperatures, from 5K to 300K. The fits are shown in grey in Fig. \ref{SI: FigSI4}.A-G. 
	
	\clearpage \subsubsection{APHE vs B} \label{SI sec: APHE vs B}
	\begin{figure*}[ht]
		\centering
		\includegraphics[width=\textwidth]{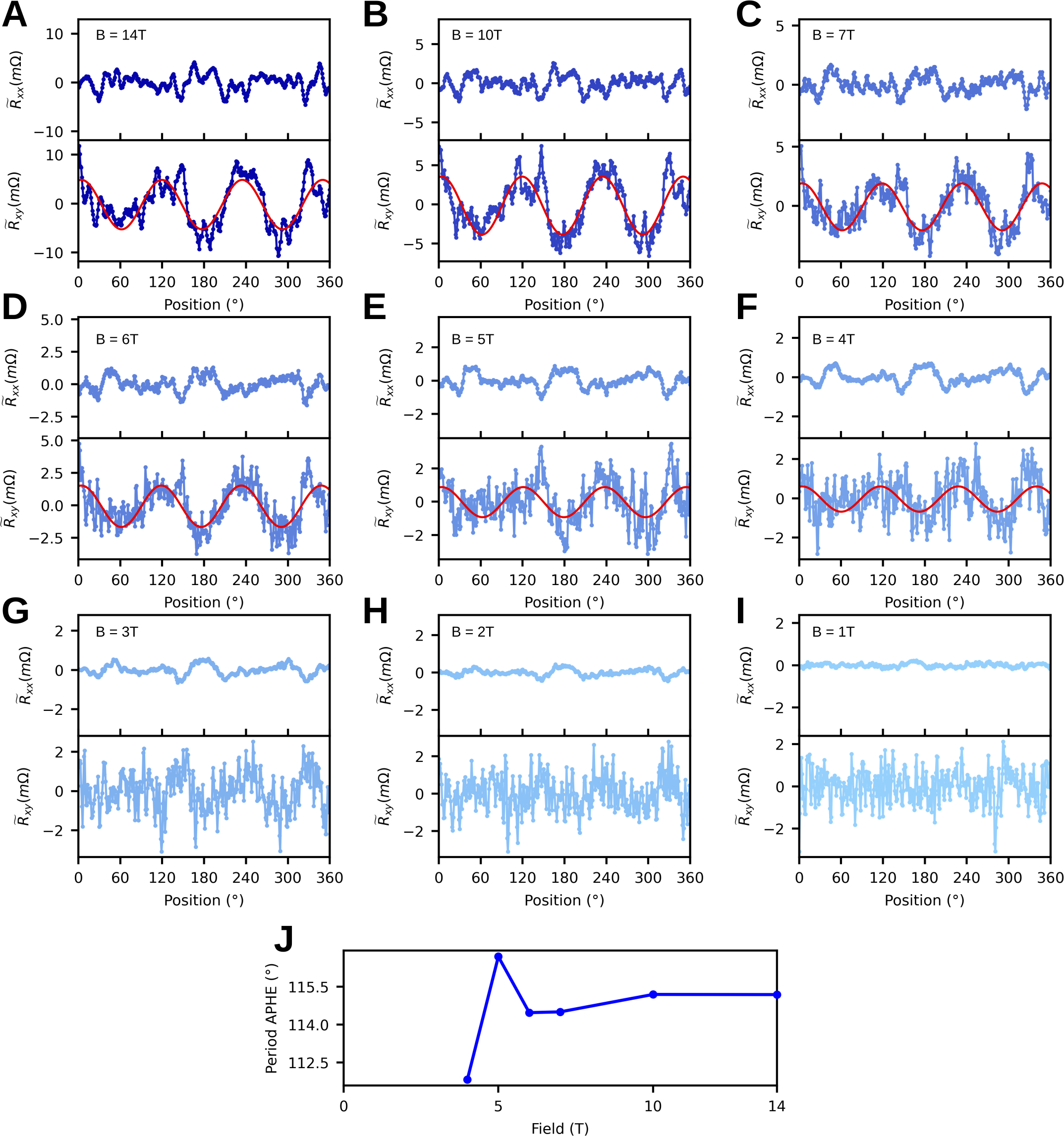}
		\caption{}
		\label{SI: FigSI5}
		\title{Residues with APHE fit at 5K, between 1T and 14T}
	\end{figure*}
	
	Fig. \ref{SI: FigSI5}.A-I shows the longitudinal and transverse residues $\Delta R = R - \overline{R}$ obtained in sec. \ref{SI sec: simplefit_B} at 5K for each magnetic field from 1T to 14T.  For $B \geq 4$~T, a roughly $2\pi/3$-periodic signal emerges in $\Delta R_{yx}$, which corresponds to the APHE. We fit this signal with the following formula:
	\begin{align} \label{SI eq: APHE_fit}
		\Delta R_{yx}(B,\varphi) &=  C(B) + A^{\text{APHE}}(B) \: \text{cos} [(\varphi-\varphi_0) \cdot  2\pi / P^{\text{APHE}}(B)], 
	\end{align}
	with $C$ a field dependent offset ; $A^{\text{APHE}}$ the amplitude of the APHE ; $\varphi_0$ the field-independent angular origin of the oscillation ; and $P^{\text{APHE}}$ the angular period of the APHE, which is considered field-dependent in the analysis. The fits obtained above 4T are shown in red in Fig. \ref{SI: FigSI5}.A-I, and fit the data very well between $120 \degree \leq \wphi \leq 360 \degree$. The residues deviate slightly from the fit below 120°, most probably due to non-uniform movement of the mechanical rotator in this angular range at low temperature. This issue is much more visible in a second sample, D2 (126nm thick, see e.g. Fig. \ref{SI: FigSI8}.A,B), and happens around specific positions in a reproducible manner. This low-angle deviation disappears at higher temperature (see sec. \ref{SI sec: APHE vs T}), which also points towards a low-temperature mechanical issue of the rotator. \\
	Fig. \ref{SI: FigSI5}.J shows the field dependence of $P^{\text{APHE}}$ yielded by the fit. Above 6T, it is stable at $P^{\text{APHE}} \sim 115$°, or just under the 120° expected for the PHE from theory. The low-field variation of the period could be the result of lower signal/noise ratio combined with the mechanical rotation issues. The variation of $A^{\text{APHE}}$ with field is shown in the main text in Fig. \ref{Fig2_Experiment}.g.

	\clearpage \subsubsection{APHE vs T} \label{SI sec: APHE vs T}
	\begin{figure*}[ht]
		\centering
		\includegraphics[width=\textwidth]{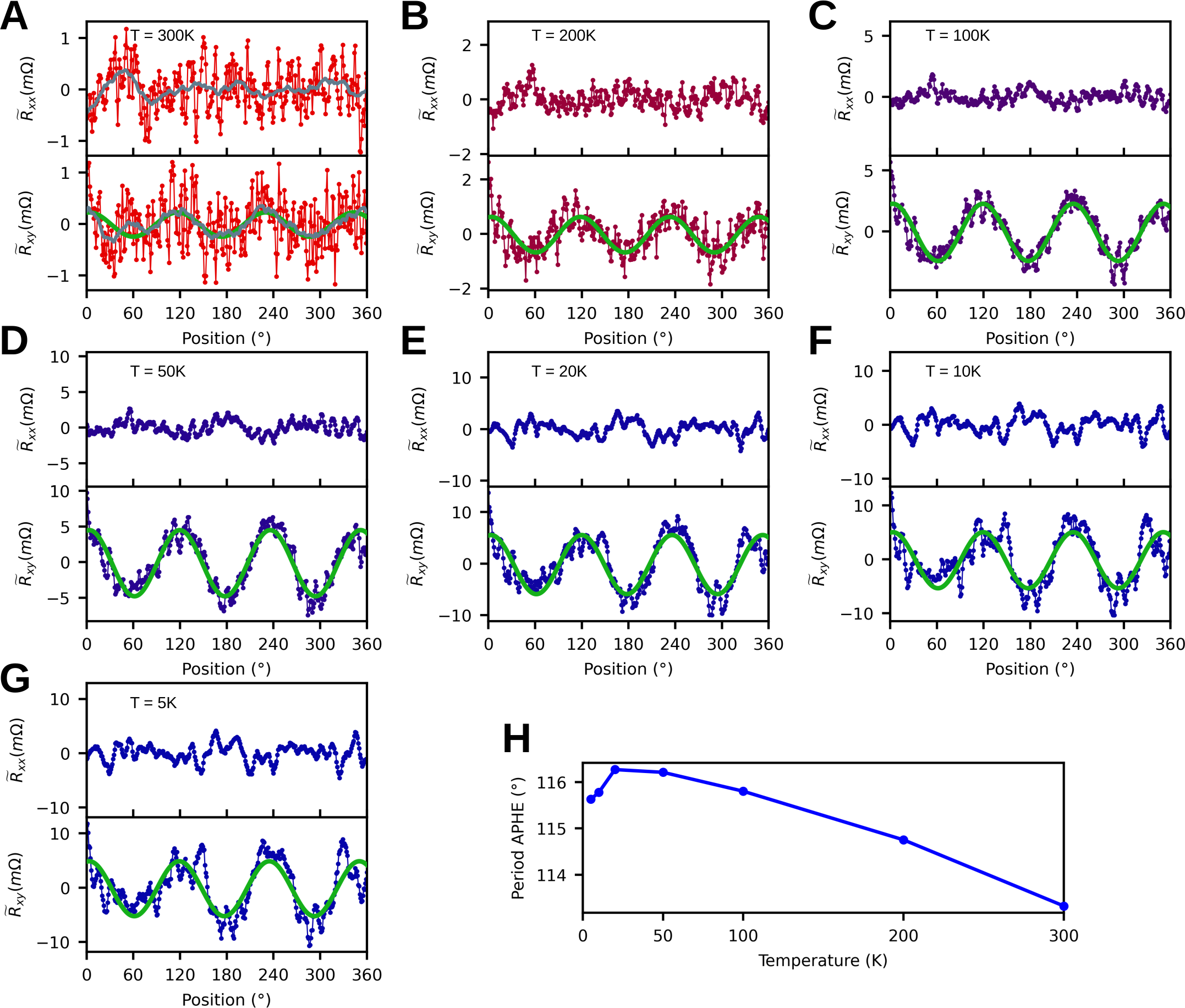}
		\caption{}
		\label{SI: FigSI6}
		\title{Residues with APHE fit at 14T, between 5K and 300K}
	\end{figure*}
	
	Fig. \ref{SI: FigSI6}.A-G shows the longitudinal and transverse residues $\Delta R = R - \overline{R}$ obtained in sec. \ref{SI sec: simplefit_T} at 14T for each temperature from 5K to 300K (same data as in Fig. \ref{Fig2_Experiment}.d-f, main text). The raw residues are fitted with eq. S6, with the temperature as a parameter instead of the field, and the resulting fits are shown in green. The low-angle deviation from the fit observed in sec. \ref{SI sec: APHE vs B} persist until $ T \sim 50$~K, above which they are no longer visible, and the fit follows the data closely over the entire angular range. \\
	As the temperature is increased, the amplitude of the APHE decreases (see Fig. \ref{Fig2_Experiment}.h, main text). At 300K (Fig. \ref{SI: FigSI6}.A), the APHE is no longer visible to the naked eye because of the low signal/noise ratio. However, by applying a simple Savitzky-Golay smooth (in grey), the $2\pi/3$-periodic oscillation becomes clearly visible, and follows closely the fit of the raw data (in green). Applying the same smooth to the longitudinal residuals does not yield any similar trend. \\
	The Savitzky-Golay smooths were done with the savgol\_filter function from the scipy.signal library in Python, with 31° of window size and a polynome order of 1. The boundary conditions were set to 'wrap' (i.e. periodic) for the transverse residues, and 'nearest' for the longitudinal residues.
	
	\clearpage \subsubsection{PHE and APHE at 126nm} \label{SI sec: 126nm}
	\begin{figure*}[ht]
		\centering
		\includegraphics[width=\textwidth]{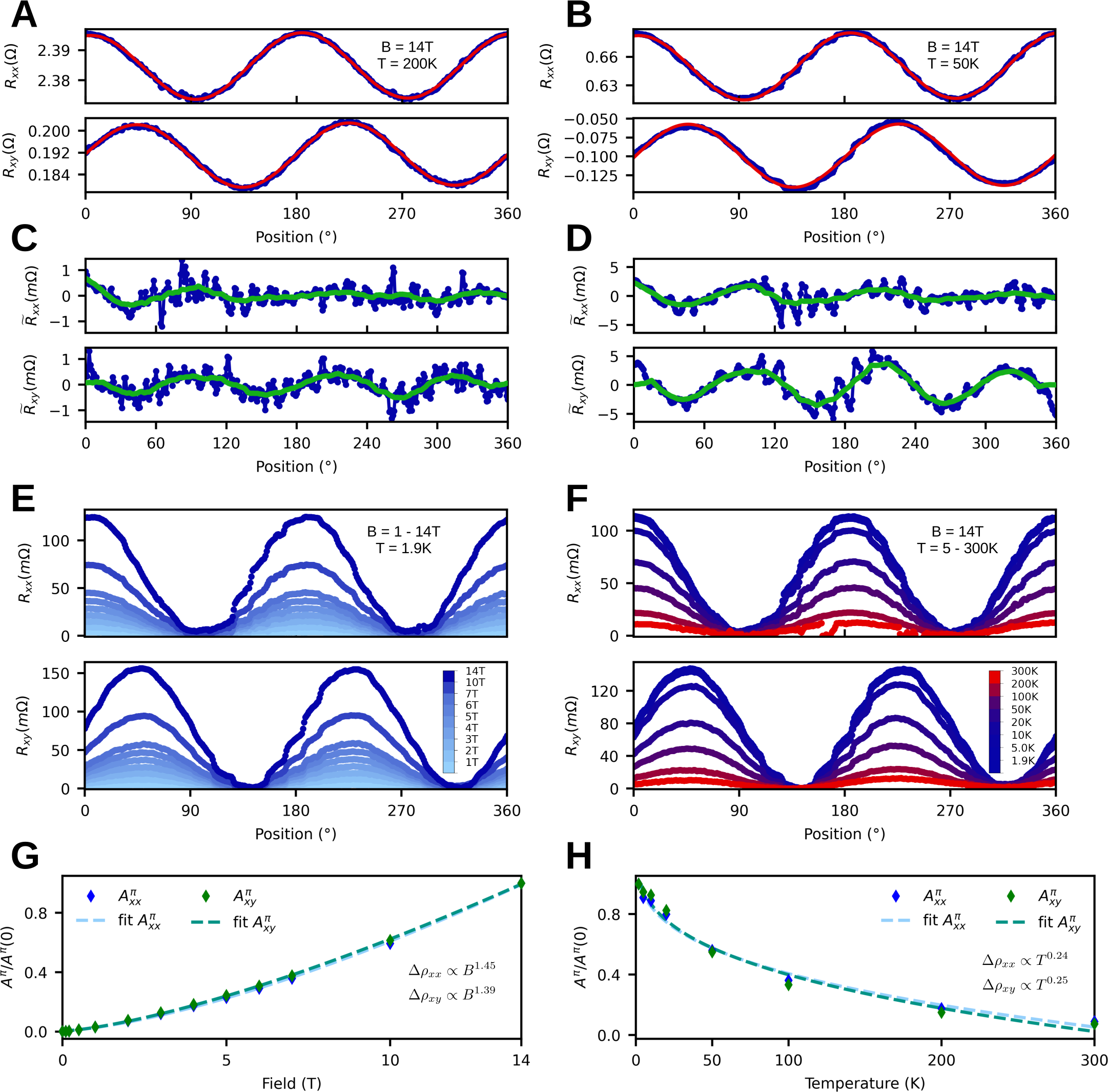}
		\caption{}
		\label{SI: FigSI8}
		\title{APHE for 126nm sample}
	\end{figure*}
	
	Similar measurements and analysis as for sample $D1$ were performed on another sample, $D2$ (126nm thick). However, this sample proved more sensitive to the issues of the mechanical rotator than D1, and the APHE signal was not necessarily visible in all residues extracted, due to increased noise levels. \\
	Fig. \ref{SI: FigSI8} shows the main results for sample D2. Fig. \ref{SI: FigSI8}.A,C show the resistance and residues at 200K, 14T. The resistance is fitted with eq. S5 (red), and the residues are smoothed with the same procedure as described in sec. \ref{SI sec: APHE vs T} (green). Fig. \ref{SI: FigSI8}.B,D show the same data, but at 50K, 14T. At both temperatures, the data residues show a clear oscillation corresponding to the APHE. \\
	
	\clearpage \subsubsection{PHE and APHE at 320nm} \label{SI sec: 320nm}
	\begin{figure*}[ht]
		\centering
		\includegraphics[width=\textwidth]{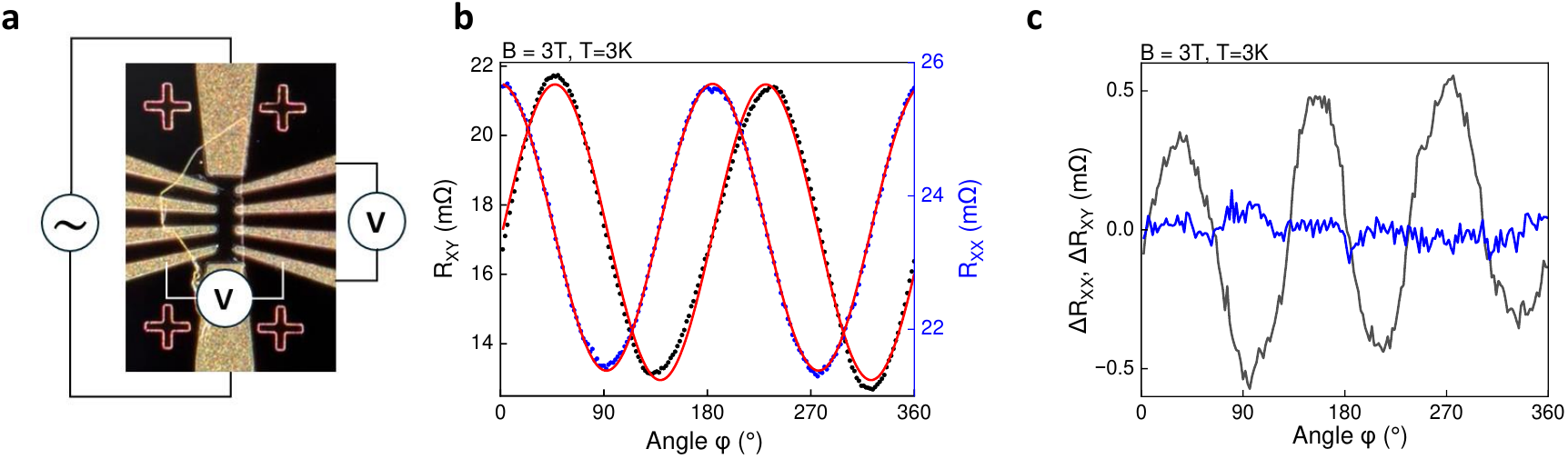}
		\caption{ \textbf{a:} Optical image of the device and schematic diagram of the measurement configurations. The Hall bar is 50$\mu$m long and 10$\mu$m large.\textbf{b}: Angular dependence of $R_\text{yx}$ (black) and $R_\text{XX}$ (blue) at $B=3$~T, and $T=3$~K. The best fits with \ref{eq: PHE} (Methods) are superimposed in red. \textbf{c}: Angular dependence of the residues $\Delta R_\text{yx}$ (grey) and $\Delta R_\text{yx}$ (blue) after a background removal. Only a $\pi$-periodic signal was removed for the present analysis and a 2$\pi$-periodic signal is therefore still visible in the residues of $\Delta R_\text{yx}$.}
		\label{SI: FigSI8a}
		\title{Standard and Anomalous Planar Hall effect in a 320 nm thick \ce{PtBi2} flake}
	\end{figure*}
	
	We have also exfoliated and measured a 320nm thick \ce{PtBi2} flake (sample D3, see fig.~\ref{SI: FigSI8a}~a).  The flake was contacted by optical lithography and for such thick flake, the residual resistance ratio is much larger than for D1 or D2 : $RRR=25$.
	
	As already stated in the methods, the flake was measured in a different set-up and using a 3D-piezorotator. The analysis was done following equation \ref{eq: PHE} in the Methods, and only a $\pi$-periodic signal was removed from the longitudinal and transverse resistances (fig.~\ref{SI: FigSI8a}~b). As a result, an additional weak 2$\pi$ signal can be seen in the residual of the transverse resistance. As a confirmation of the results obtained for D1 and D2, a very clear $2\pi/3$-periodic signal can be seen in the transverse resistance whereas no $2\pi/3$-periodic oscillation could be identified in the longitudinal resistance (fig.~\ref{SI: FigSI8a}~c). 
	
	\begin{figure*}[ht]
		\centering
		\includegraphics[width=0.5\textwidth]{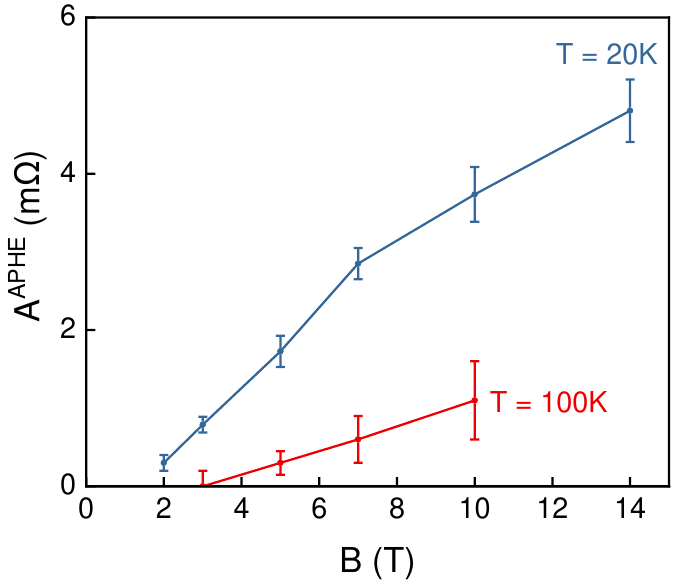}
		\caption{Amplitude of the APHE signal as a function of the magnetic field and for $T=20$~K (blue) and $T=100$~K. }
		\label{SI: FigSI8b}
		\title{APHE versus magnetic field at $T=20$~K and 100~K}
	\end{figure*}
	
	For D3, the evolution of the APHE signal was measured as a function of the magnetic field for two temperatures : $T=20$~K and $T=100$~K (fig.~\ref{SI: FigSI8b}). A clear temperature dependence of the field onset could be evidenced, with a higher onset seen at higher temperature, supporting our hypothesis that the field onset is linked to the effective gap of the TNLs.\\

	\clearpage \subsubsection{Residual Resistance Ratio} \label{SI sec: RRR}
	\begin{figure*}[ht]
		\centering
		\includegraphics[width=\textwidth]{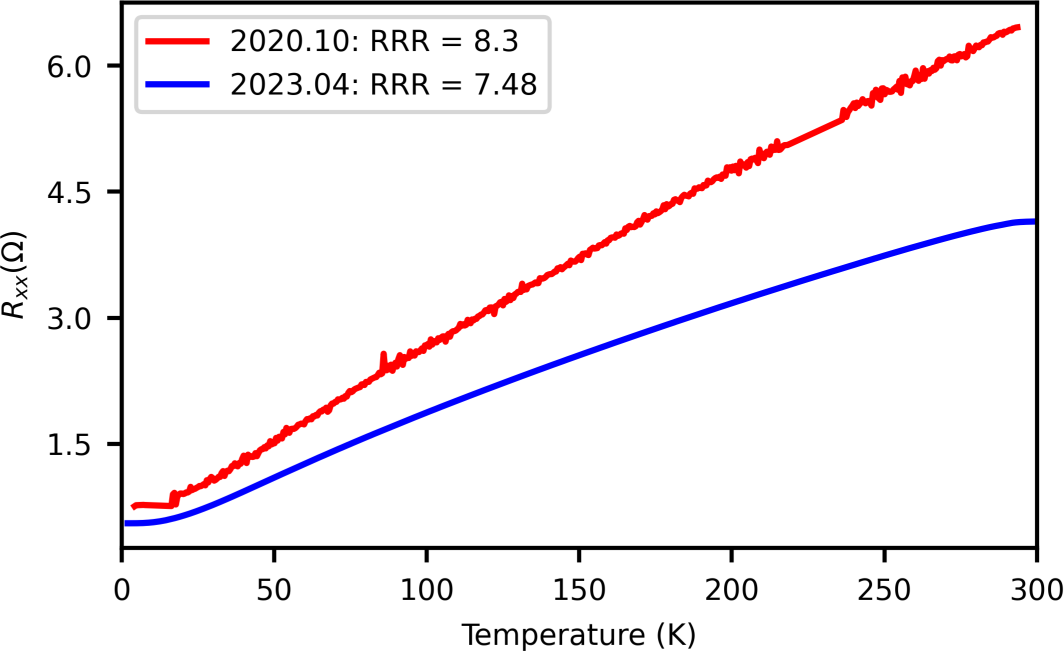}
		\caption{}
		\label{SI: FigSI9}
		\title{RRR for D1 between 2020 and 2023}
	\end{figure*}
	
	Fig. \ref{SI: FigSI9} shows a comparison of the longitudinal resistance of sample $D1$ between two cooldowns from room temperature to 5K, in 2020 (red, see Ref.\cite{Veyrat2023}) and in 2023 (blue, this measurement run). The residual resistance ration $RRR = \dfrac{R(300K)}{R(5K)}$ is similar between the two cooldowns, and only decreased slightly in 2.5 years, showing the stability of the samples in air.

	\clearpage
	
	\subsubsection{The Weyl point geometry of band 48\label{subsec:suplmat-Weyl-point-geometry}}
	
	We focus on the band number 48, which when being the highest fully occupied
	band yields the correct number of electrons of the system. Of course the 
	compound is not an insulator, however the Weyl points reported before 
	are formed by band 48 and 49 and in the region $k_z\lesssim 0.17 \frac{2\pi}{c}$
	the highest occupied band is band 48.
	
\begin{figure}[htpb]
		\begin{centering}
			\includegraphics[width=0.8\columnwidth]{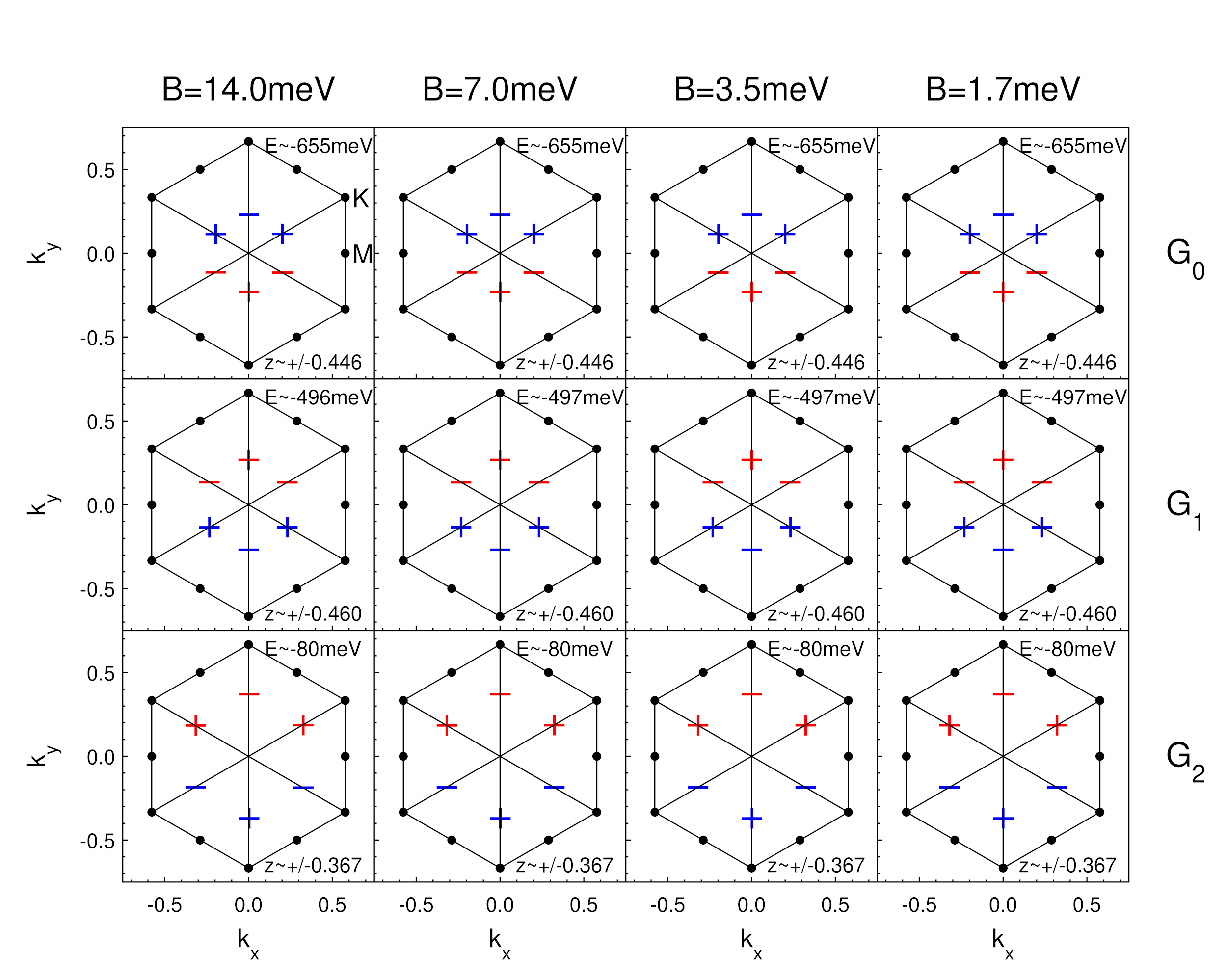}
			\par\end{centering}
		\caption{\label{fig:WP-groups-012}WP groups 0, 1 and 2 of band 48. The average energy
			and $k_{z}$-position is denoted. Plus/minus signs denote the sign
			of the $k_{z}$-position and red/blue mark positive/negative chirality,
			resp.}
	\end{figure}

\begin{figure}[htbp]
		\begin{centering}
			\includegraphics[width=0.8\columnwidth]{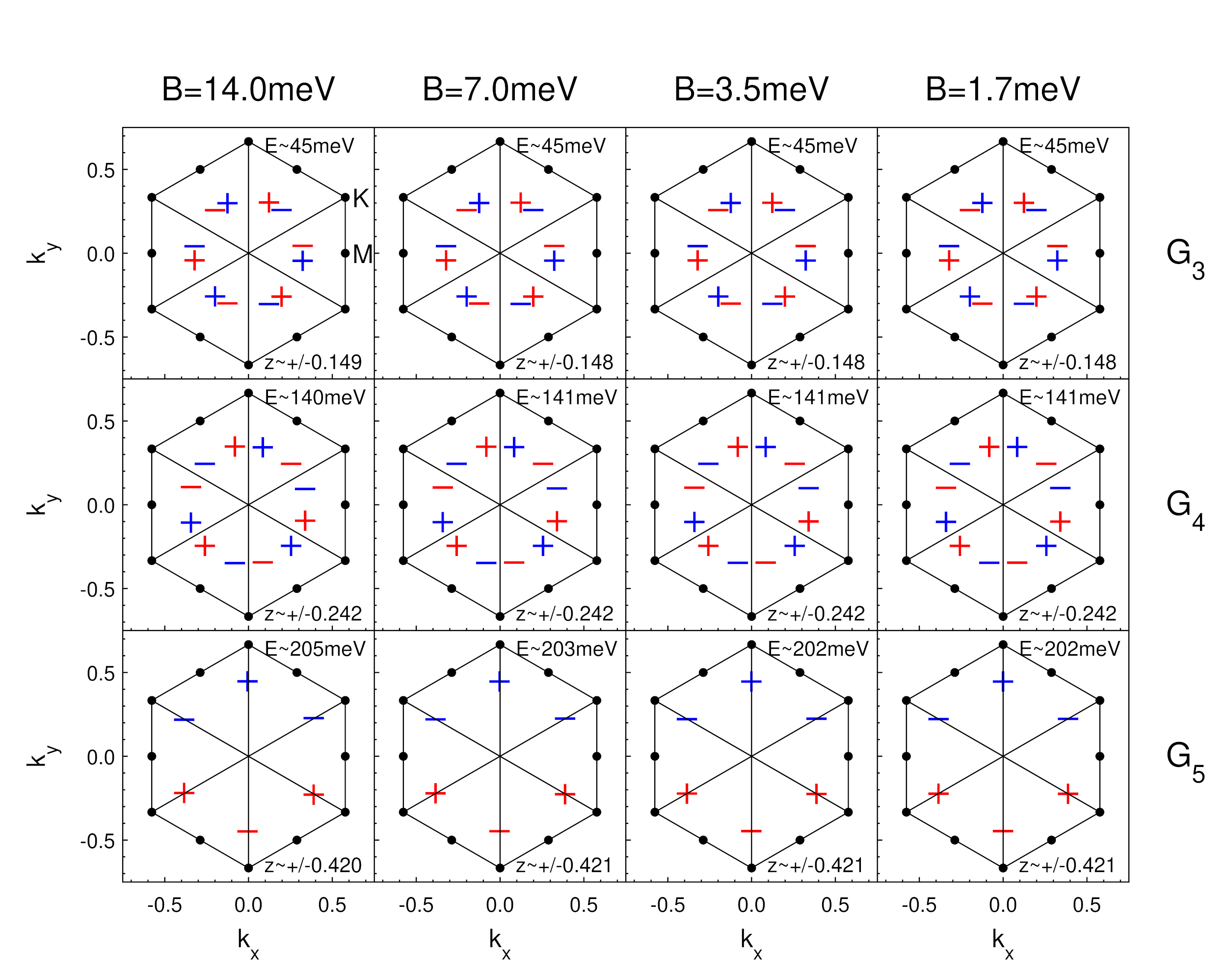}
			\par\end{centering}
		\caption{\label{fig:WP-groups-345}WP groups 3, 4 and 5 of band 48. Group 3 is the one
			close to the Fermi level. The average energy and $k_{z}$-position
			is denoted. Plus/minus signs denote the sign of the $k_{z}$-position
			and red/blue mark positive/negative chirality, resp.}
	\end{figure}

\begin{figure}[htbp]
		\begin{centering}
			\includegraphics[width=0.8\columnwidth]{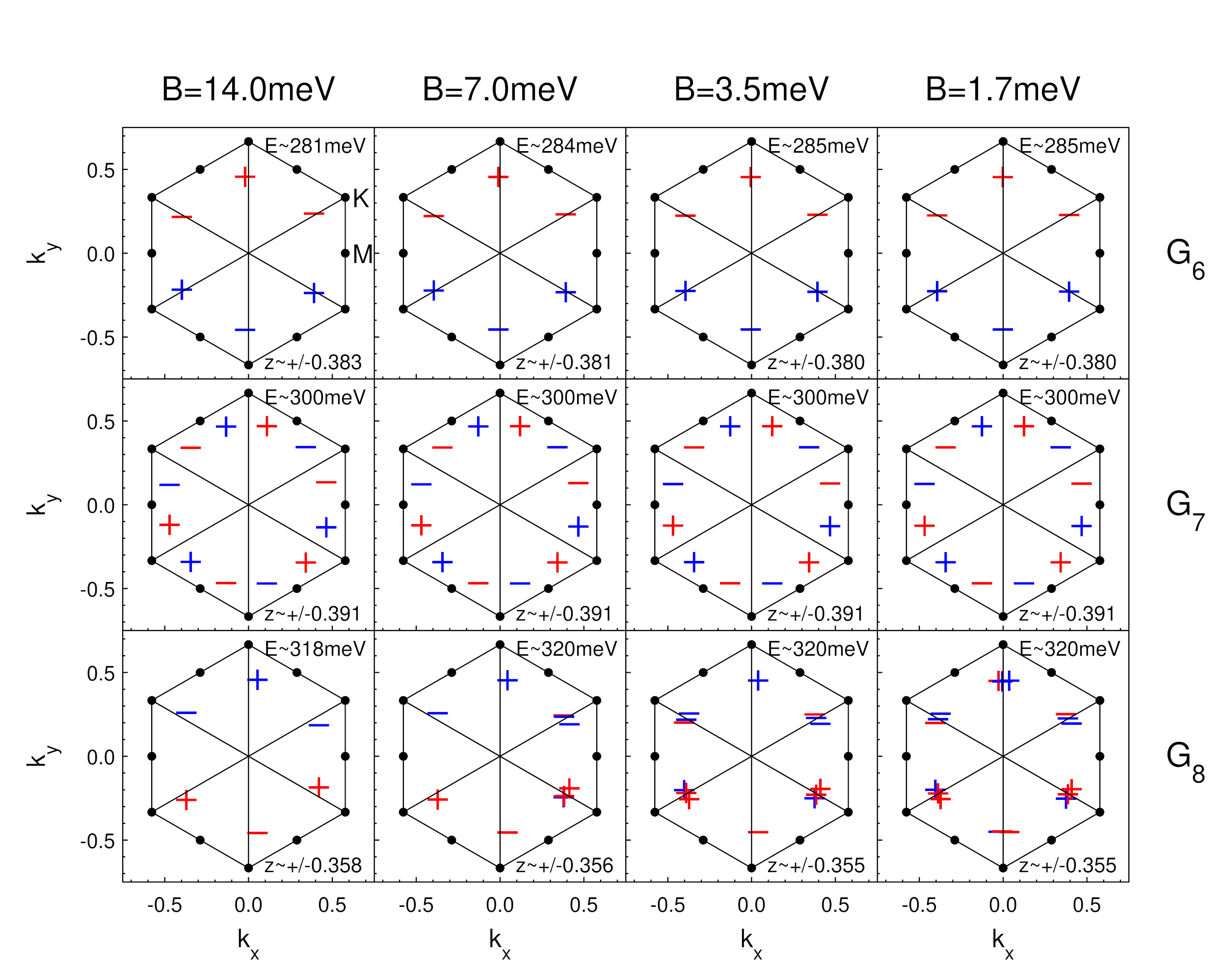}
			\par\end{centering}
		\caption{\label{fig:WP-groups-678}WP groups 6, 7 and 8 of band 48. The average energy
			and $k_{z}$-position is denoted. Plus/minus signs denote the sign
			of the $k_{z}$-position and red/blue mark positive/negative chirality,
			resp.}
	\end{figure}

	As it turns out band 48 has more than the previously reported Weyl points.
	We group the Weyl points (WP) according to their averaged energy.
	We get six groups of six and three groups of twelve WPs with the exception
	of group $\mathrm{G}_{8}^{48}$, see below. Within each group the WPs have
	individual $k_{x,y,z}$ coordinates of which the $k_{z}$ positions
	are clustered very closely around $\pm\left \vert k_{z0}\right \vert$ with a
	group specific $\left\vert k_{z0}\right \vert$. The only (always fulfilled)
	true symmetry is $\Theta M_{x}$. However, the deviations from the
	non-magnetic symmetries are small, such that the twelve-fold groups
	approximately follow $\left\{ E,\Theta\right\} C_{3v}$, while the
	six-fold groups follow this symmetry if the WPs are assumed to lie
	exactly on the mirror planes. Group $\mathrm{G}_{8}^{48}$ is an exception
	in that it shows cross-over behavior between a 12-fold and a six-fold
	case as a function of increasing magnetic field. From the symmetry
	and from the small displacements shown in Fig. \ref{fig:WP-groups-012},
	Fig. \ref{fig:WP-groups-345} and Fig. \ref{fig:WP-groups-678} it is clear
	that the twelve-fold groups survive the transition to $B=0$, while
	the six-fold groups must dissolve in the nodal loops on the mirror
	planes. 
	
	\begin{figure}[htbp]
		\begin{centering}
			\includegraphics[width=0.6\columnwidth]{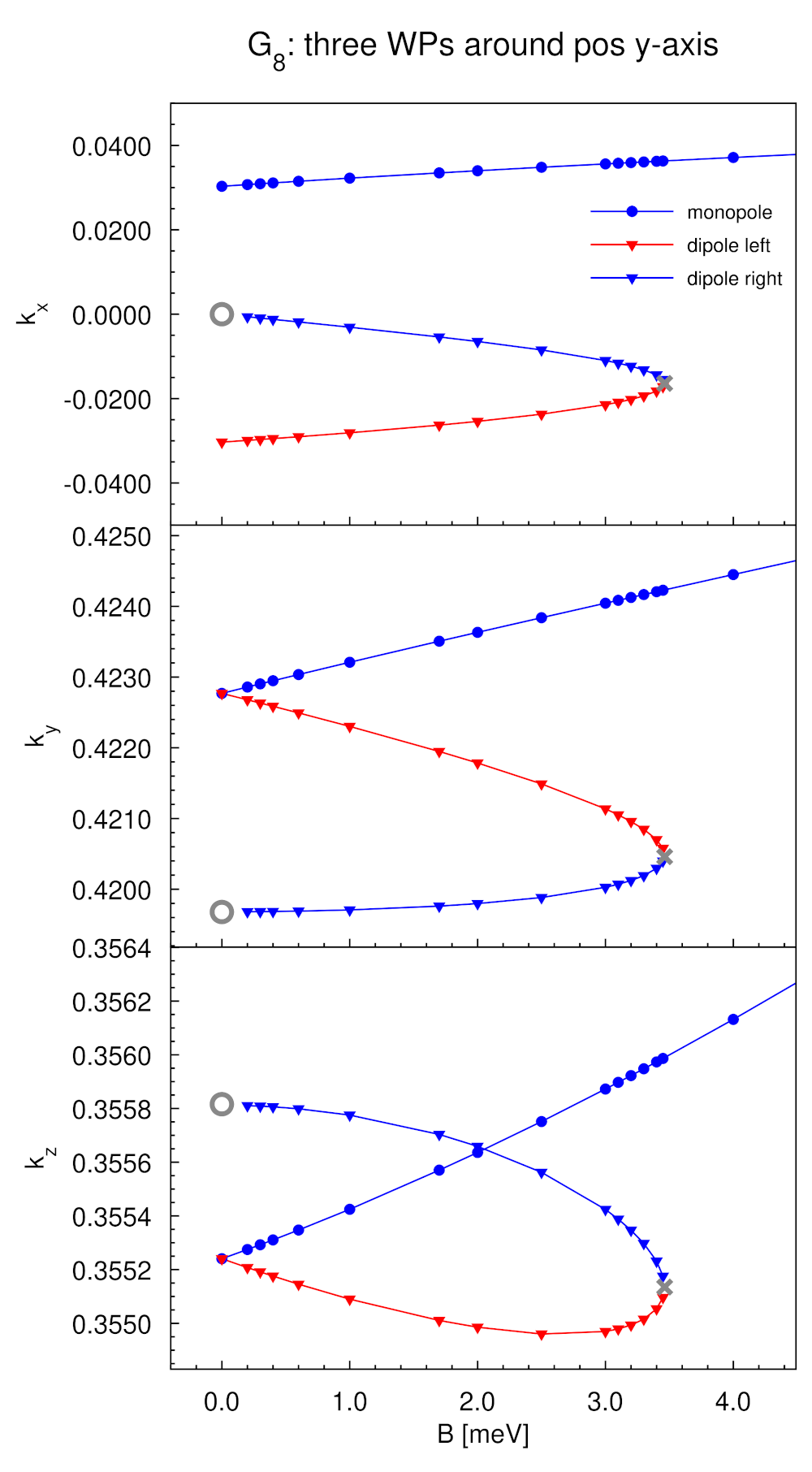}
			\par\end{centering}
		\caption{\label{fig:evolutiongroup8}Evolution of the three WPs close to the
			positive $y$-axis of $\mathrm{G}_{8}^{48}$ for small magnetic fields. The open
			circle denotes the position at which one of the WPs dissolves into
			the nodal loop. At $B=0.35$ a dipole emerges at negative $k_{x}$
			(gray cross). Compare to last row of Fig. \ref{fig:WP-groups-678}.}
	\end{figure}
	
	The cross-over behavior of group 8 demonstrates both the survival
	of the twelve-fold groups as well as the dissolution of the six-fold
	groups. In Fig. \ref{fig:evolutiongroup8} we follow the evolution
	of the three WPs close to the positive $y$-axis of Group $\mathrm{G}_{8}^{48}$
	(Fig. \ref{fig:WP-groups-678} lower right panel). For high field
	($B=14\mathrm{meV}$) we have a six-fold group (in this case shifted
	to the positive $k_{x}$-side of the mirror). At around $B=3.5\mathrm{meV}$
	a WP dipole emerges on the negative $k_{x}$-side of the mirror, whose
	separation increases with decreasing field (Fig. \ref{fig:evolutiongroup8}).
	Approaching $B=0$ the dipole part with equal chirality as the single
	WP from the six-fold subgroup is dissolved into the nodal loop, while
	the dipole part with opposite chirality to the single WP approaches
	the mirror symmetric position of the single six-fold WP, forming a
	fully symmetric twelve-fold group for $B=0$. Apparently the emergence
	of the WP dipoles happens at different fields for the six points of
	the high-field six-fold group (last row of Fig. \ref{fig:WP-groups-678})
	as allowed by the magnetic symmetry.
	We want to stress that there is no symmetry which would force the appearance 
	of six-fold or twelve-fold groups as already indicated by the behaviour of $\mathrm{G}_{8}^{48}$.
	A group could have any even number of points as long as they occur in pairs as 
	imposed by the magnetic symmetry.
	
	It turns out that all but groups $\mathrm{G}_{3,4}^{48}$ consist of at least some type II 
	Weyl points, while these two groups are formed only by type I points.

	\subsubsection{Chern signal}\label{SI subsec:Chern-signal}
	
	The Chern number of a $k_{z}$-plane is defined as 
	\begin{equation}
		c\left(k_{z}\right)=\frac{1}{2\pi}\int dk_{x}dk_{y}F_{z}\left(\boldsymbol{k}\right)\label{SI eq: Chernnumber}
	\end{equation}
	with $\boldsymbol{A}\left(\boldsymbol{k}\right)=\sum_{n}\left\langle u_{n}\mid i\nabla_{\boldsymbol{k}}u_{n}\right\rangle $,
	$\boldsymbol{F}\left(\boldsymbol{k}\right)=\nabla_{\boldsymbol{k}}\times\boldsymbol{A}\left(\boldsymbol{k}\right)$,
	while the total anomalous Hall conductivity (AHC) is defined as 
	\begin{align*}
		\Delta\sigma_{xy} & =-\frac{e^{2}}{\hbar}\frac{1}{\left(2\pi\right)^{3}}\int d^{3}kF_{z}\left(\boldsymbol{k}\right)\\
		& =-\frac{e^{2}}{\hbar}\frac{1}{\left(2\pi\right)^{2}}\int dk_{z}c\left(k_{z}\right)\\
		& =-\frac{e^{2}}{\hbar}\frac{1}{c2\pi}\int_{0}^{1}dzc\left(z\right)
	\end{align*}
	with $c=6.162\text{Å}$ and conversion constant 
	\begin{equation}
		\frac{e^{2}}{\hbar}\frac{1}{c2\pi}  =628.71\mathrm{S/cm}\label{eq:conv_factor}
	\end{equation}
	The Chern signal $c\left(k_{z}\right)$ can be obtained by a gauge
	invariant plaquette type integral \cite{Fukui2005}, which has better
	convergence properties than a simple sum-type integral, but is only
	valid for a constant number of occupied bands, i.e. an insulating
	$k_{x},k_{y}$-plane. $c\left(k_{z}\right)$ is given by the sum of
	$\tilde{c}=\frac{1}{2\pi}i\ln U_{1}U_{2}U_{3}U_{4}$ over all plaquettes
	of the planar Brillouin zone perpendicular to $k_{z}$ with $U_{i}=\frac{\mathrm{det}\left\langle u_{k_{i}}\middle \vert u_{k_{i+1}}\right\rangle }{\left\vert\mathrm{det}\left\langle u_{k_{i}}\middle \vert u_{k_{i+1}}\right\rangle \right\vert}$,
	where $k_{i}$ are the $k$-points at the four plaquette corners.
	The determinant is taken of the overlap matrix formed by the 48 (homo)
	lowest bands. The minimally needed mesh fineness is dictated by the
	WP separation in the $k_{x},k_{y}$ plane and the mesh fineness of
	the $k_{z}$-interval by the $k_{z}$-separation of the individual
	WPs, which turns out to be demanding. 
	Additionally, the experimentally
	used magnetic fields ($\boldsymbol{B}$) are small from a band structure
	perspective and the subsequent WP shifts (and emergence of new WPs)
	leads to strong variations of $\boldsymbol{F}$ over rather small
	volumes.
	
	To incorporate a Fermi energy eq. S7 is evaluated via
	a simple Riemann sum.
	
\begin{figure}[htbp]
		\begin{centering}
			\includegraphics[width=1.0\columnwidth]{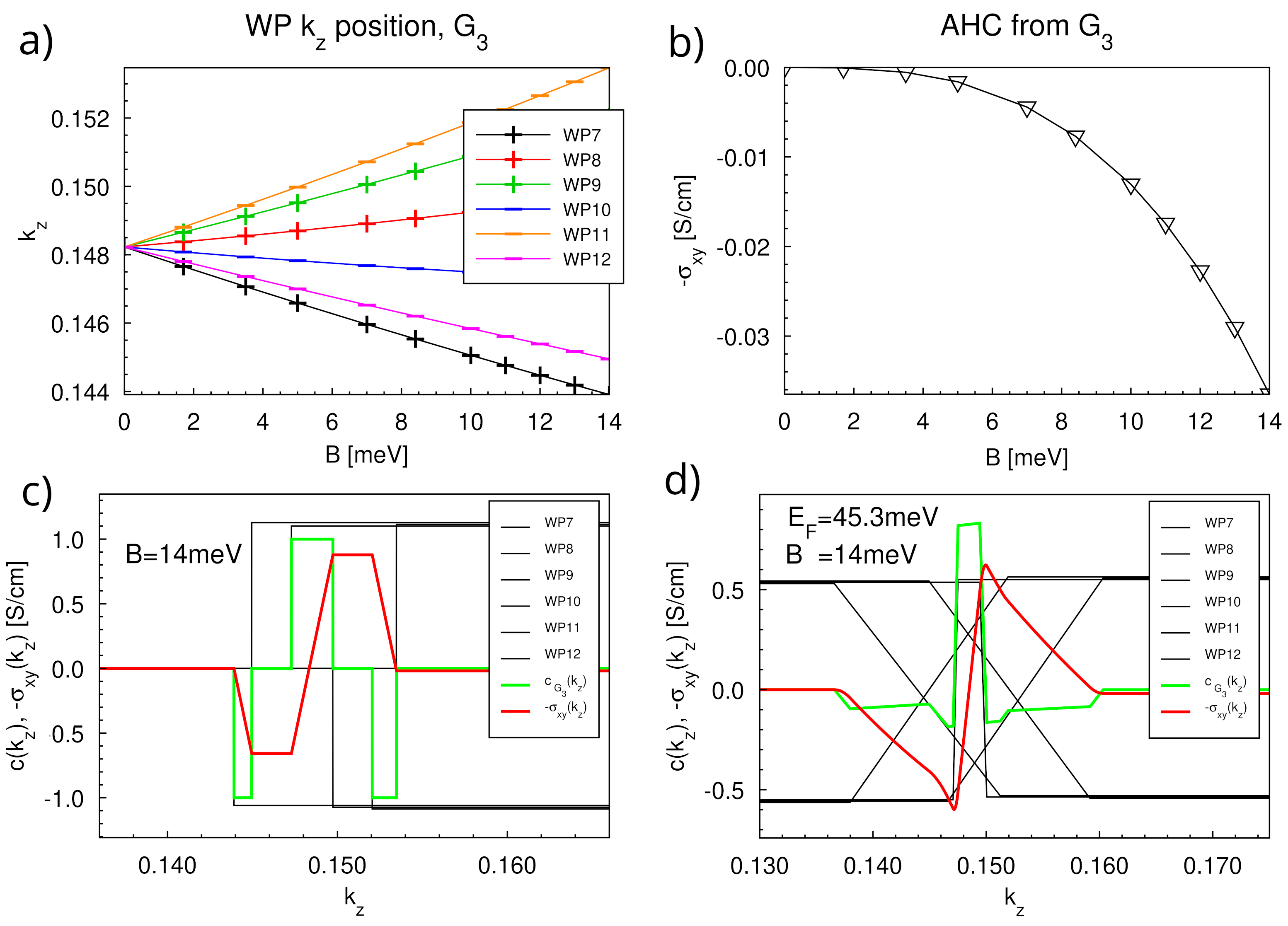}
		\end{centering}
		\caption{\label{fig:The-AHC-signal-G3}a) the evolution of the WP $k_{z}$-position
			of the six WP with $k_{z}>0$ of $\mathrm{G}_{3}^{48}$ as a function of
			$B$. The plus/minus symbols denote the chirality. b) The resulting
			AHC signal for the fully linearly compensated group $\mathrm{G}_{3}^{48}$
			as a function of field $B$ assuming ideal Weyl dispersion. c) The
			individual contributions (jumps) due to the six $k_{z}>0$ WPs of
			$\mathrm{G}_{3}^{48}$ (black) offset along the $y$-axis for better visibility,
			their sum $c\left(k_{z}\right)$ and its $k_{z}$-integral $-\Delta\sigma_{xy}\left(k_{z}\right)$
			in S/cm (red). d) As c) but for $E_{\mathrm{F}}=45.3\mathrm{meV}$. }
	\end{figure}

	The Chern signal for the insulating case is entirely determined by
	the Weyl positions and hence the numerical results can be checked
	analytically. Each WP contributes a jump 
	\[
	c_{\mathrm{WP}}\left(k_{z}\right)=\chi_{\mathrm{WP}}\Theta\left(k_{z}-k_{z,\mathrm{WP}}\right)
	\]
	The integral over $k_{z}$ gives the contribution $\Delta\sigma_{xy,\mathrm{WP}}\left(k_{z}\right)\propto-\chi_{\mathrm{WP}}\left(k_{z}-k_{z,\mathrm{WP}}\right)\Theta\left(k_{z}-k_{z,\mathrm{WP}}\right)$.
	The sum over all Weyl points for large $k_{z}$ (total signal) then
	is 
	\begin{equation}
		\Delta\sigma_{xy}\propto\sum_{\mathrm{WP}}\chi_{\mathrm{WP}}k_{z,\mathrm{WP}}\label{eq:sigma_from_kz}
	\end{equation}
	For a subgroup of six WPs with vanishing total chirality around some
	positive $k_{z}$ (half of a twelve-fold group) $\Delta\sigma_{xy}$ becomes
	a function of the $k_{z}$-separation of the individual WPs. It turns
	out that the WPs are ordered in three pairs of two points with opposite
	$\chi$. Hence, each pair contributes an up-down jump combination
	(Fig. \ref{fig:The-AHC-signal-G3}.C) with $\Delta\sigma=\chi\Delta k_{z}$
	which is proportional to the separation of the points in the pair
	(up to some global integer, which is determined by the numerical integrals).
	The separations $\Delta k_{z}$ are small and in leading order a linear
	function of the field $B$ (Fig. \ref{fig:The-AHC-signal-G3}.A).
	Additionally, for group $\mathrm{G}_{3}^{48}$, the separations $\Delta k_{z}$
	together with their $\chi$ compensate the linear field dependence
	of $\Delta k_{z}$ completely, which results in very tiny total $\Delta\sigma_{xy}$
	values, which depend quadratic on $B$ (Fig. \ref{fig:The-AHC-signal-G3}.B). 
	
	The $\chi$-uncompensated three WP subgroups at positive $k_{z}$
	of a six-fold group give an effective jump $\Delta c=\pm1$ at the
	group's averaged $k_{z}$-position. Since, the $k_{z}$ positions
	of the groups can have sizable distances these well separated jumps
	contribute large values to $\Delta\sigma_{xy}$, in fact much larger than
	the actual result including the Fermi surface (black line)
	in Fig. \ref{Figure3_Theory} of the main text and
	Fig. \ref{fig:resolved-Chern-homo48-homo47}
	(keep in mind the conversion factor eq. S8).
	
	\begin{figure}[htbp]
		\begin{centering}
			\includegraphics[width=0.8\columnwidth]{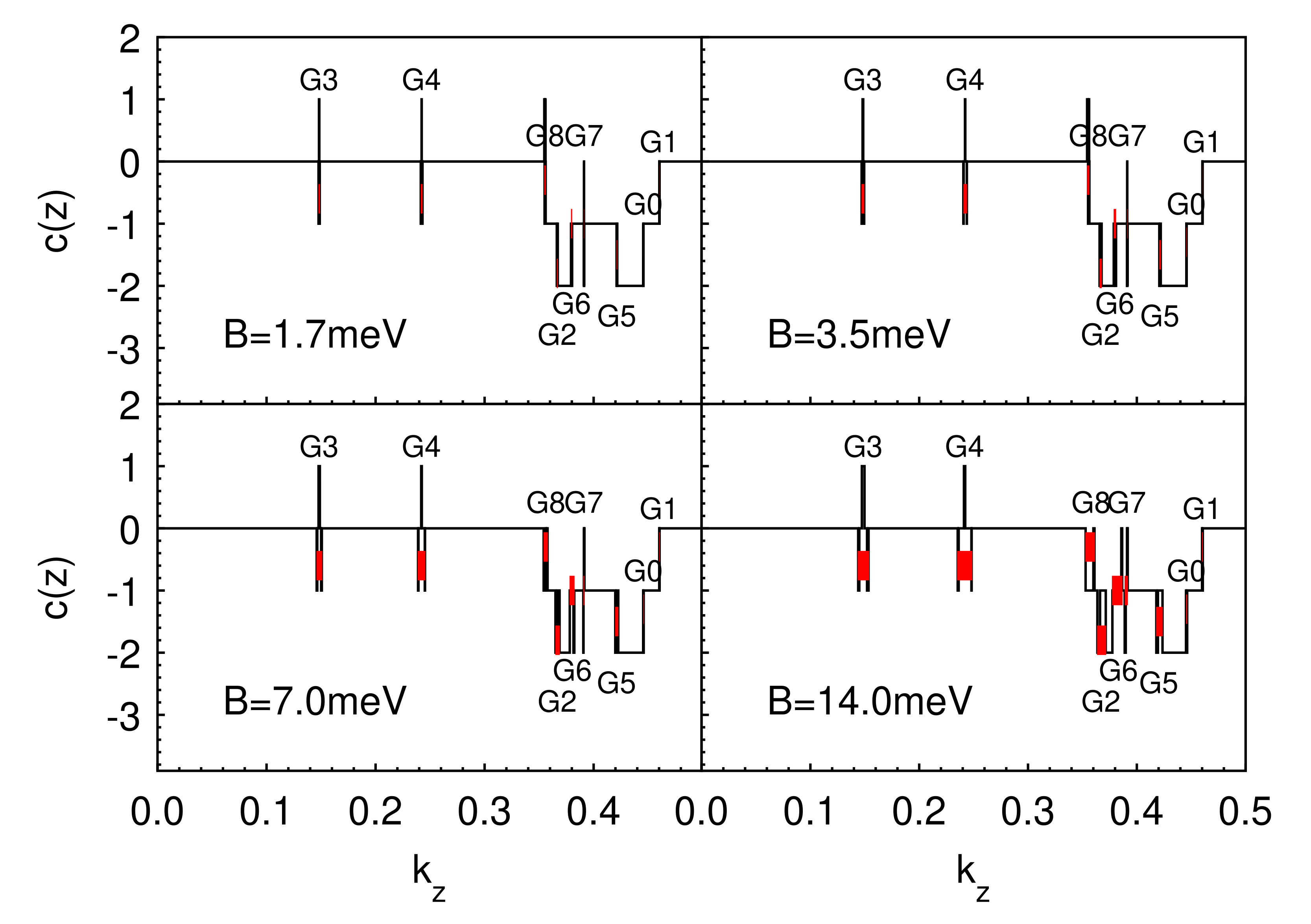}
			\par\end{centering}
		\caption{\label{fig:integer_contrib_four_B}Chern signal $c\left(k_{z}\right)$
			for four magnetic fields for constant homo 48. The jumps are labeled by the corresponding
			WP groups. The red bars span all WPs of the corresponding group.}
	\end{figure}
	
	The total result for $c\left(k_{z}\right)$ for four different fields
	$B$ is shown in Fig. \ref{fig:integer_contrib_four_B}. Clearly
	the twelve-fold groups $\mathrm{G}_{3,4,7}^{48}$ have compensating jumps within
	themselves and the separation of the individual jumps within such
	a group diminishes with diminishing field (Fig. \ref{fig:The-AHC-signal-G3}.A).
	Each of these groups contributes a signal similar to the green line
	in Fig. \ref{fig:The-AHC-signal-G3}.C. In contrast, the six-fold
	groups contribute effective jumps $\Delta c=\pm1$ at different $k_{z}$
	positions each consisting of three closely spaced up-down-up or down-up-down
	jumps, whose internal separation diminishes with $B$, effectively
	becoming a single jump. Since the $k_{z}$ positions of the group
	do not move around much with field, different groups do not compensate
	each other for $B\to0$. The corresponding signal comes into existence
	immediately when $B>0$, where these groups emerge out of the nodal
	loops. This situation is reminiscent of a quantum hall situation.
	
	The discussion up to now was for a hypothetical insulating case where
	all bands up to the homo (48) are considered occupied. This means that
	all WPs of all groups are occupied exactly up to the touching point
	of the linear dispersion. This leads to the vertical jumps shown in
	Fig. \ref{Figure3_Theory} of the main text, Fig. \ref{fig:The-AHC-signal-G3}.C and
	Fig. \ref{fig:integer_contrib_four_B}. If a Fermi level $\mu$ is introduced
	close enough to the energy of a WP (or WP group for that matter),
	such that the linear dispersion still holds, the jumps get modified:
	\[
	c_{\mathrm{WP}}\left(k_{z}\right)=\chi_{\mathrm{WP}}\Theta_{\mu}\left(k_{z}-k_{z,\mathrm{WP}}\right)
	\]
	with 
	\[
	\Theta_{\mu}\left(k_{z}\right)=\mathrm{sign}\left(k_{z}\right)\begin{cases}
		1 & \left\vert\mu\right\vert<\left\vert{}k_{z}\right\vert\\
		\frac{\left\vert{}k_{z}\right\vert}{\left\vert\mu\right\vert} & \left\vert{}k_{z}\right\vert<\left\vert\mu\right\vert
	\end{cases}
	\]
	and $\mu$ measured relative to the degenerate point and 
	$k_z$ being scaled with a suitable band velocity (omitted).
	Here we integrated
	over the whole infinite $k_{x},k_{y}$-plane. Finite integration over
	a disk of radius $R$ modifies the expression but not in an essential
	way. The modification is due to the necessary omission ($\mu<0$)
	are addition ($\mu>0$) of a disk of radius $\mu$ around the planar
	WP position and holds for $\mu\lessgtr0$. The curve is essentially
	a step function with a linear rise of width $\left\vert\mu\right\vert$ instead
	of a vertical jump. The corresponding $k_{z}$-integral becomes 
	\[
	\Delta\sigma_{xy}\left(k_{z}\right)\propto\chi_{\mathrm{WP}}\Delta\sigma_{\mu}\left(k_{z}-k_{z,\mathrm{WP}}\right)
	\]
	\begin{align*}
		\Delta\sigma_{\mu}\left(k_{z}\right) & =\begin{cases}
			0 & k_{z}<-\left\vert\mu\right\vert\\
			\frac{\left(k_{z}+\left\vert\mu\right\vert\right)^{2}}{4\left\vert\mu\right\vert} & \left\vert{}k_{z}\right\vert<\left\vert\mu\right\vert\\
			k_{z} & \left\vert\mu\right\vert<k_{z}
		\end{cases}
	\end{align*}
	which for $k_{z}$ above $k_{z,\mathrm{WP}}+\mu$ gives the same total
	contribution of a WP group as eq. S9. The result
	is altered locally around the groups $k_{z}$-position but far away
	the AHC signal is Fermi level independent (see also \cite{Burkov2014}). This
	is of course only true if the ideal linear Weyl dispersion holds and
	if the broadening due to the linear sections does not become comparable 
	to the size of the total $k_z$-interval.
	Interestingly, numerical results for Fermi levels close to a twelve-fold
	groups energy show a $c\left(k_{z}\right)$ signal of the group, which
	looks very similar to the idealized results derived here. For larger
	energy distances we see more and more deviations from the ideal behavior.
	
	Fig. \ref{fig:The-AHC-signal-G3}.D is the equivalent of Fig. \ref{fig:The-AHC-signal-G3}.C
	for $\mu=E_{\mathrm{F}}=45.3\mathrm{meV}$. Note, how the vertical
	jumps become linearly interpolated jumps. The linear regions of the
	two contributions around the middle of the $k_{z}$-axis are quite
	narrow, since these WPs have an energy close to $\mu$, while the
	other four WPs have larger deviations of their energy from $\mu$
	and hence wider linear regions. The resulting $\Delta\sigma_{xy}(k_{z})$ signal
	looks quite different from the constant homo case but its total value
	for large $k_{z}$ is Fermi level independent.
	
	In a real band structure the ideal WP linear dispersion bends into
	the real bands away from the WP position, which partially removes
	the Fermi level independence and leads to the washing out of the effects
	of the integer Chern numbers. Additionally, some of the WPs are highly
	anisotropic, which leads to further deviations from the idealized
	model discussed above. 
	
	This Fermi level independence gets also altered if the Weyl points are of type II,
	in which case the formulas above need to be modified 
	and additionally the associated Fermi pockets, especially if they are small, quickly lead to a deviation from an ideal Weyl Hamiltonian.
	
	\subsubsection{Band 47}\label{SI subsec: band47}
	
\begin{figure}[htbp]
		\begin{centering}
			\includegraphics[width=1.0\columnwidth]{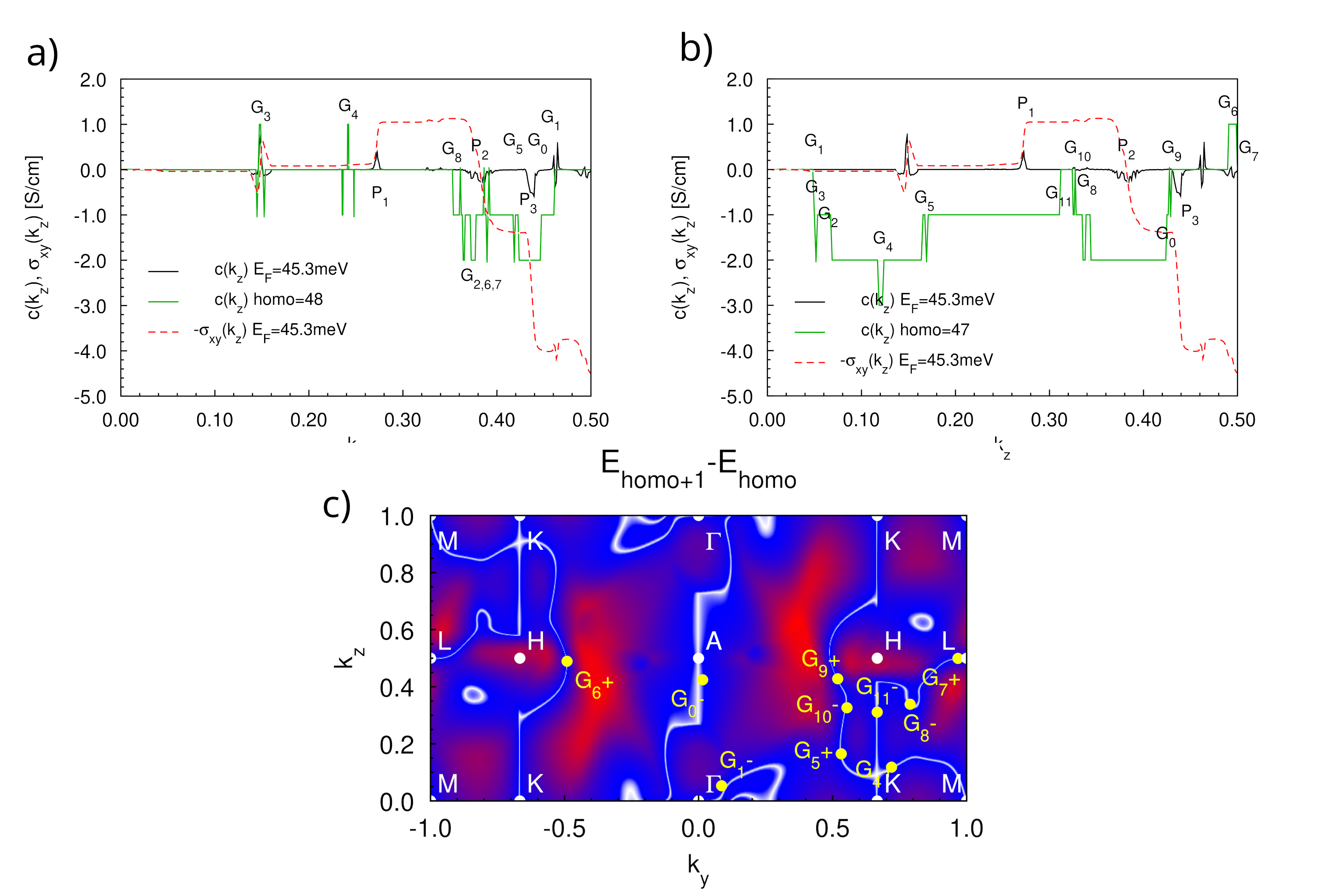}
		\end{centering}
		\caption{\label{fig:resolved-Chern-homo48-homo47} \textbf{a,b}: the $k_{z}$-resolved
			Chern number for (green) constant homo, (black) constant Fermi energy
			and (red-dashed) the resulting AHC signal $-\Delta\sigma_{xy}\left(k_{z}\right)$
			for \textbf{a}: homo 48 and \textbf{b}: homo 47. \textbf{c}: the Energy gap for homo 47. White
			denotes zero gap (nodal lines). All groups except $\mathrm{G}_{2,3}^{47}$ are shown
			and fall close the nodel lines. $\mathrm{G}_{2,3}^{47}$ are two-point groups and
			are closer to other mirror planes and hence do not fit into this picture.}
	\end{figure}

	Unsurprisingly, the magnetic field also induces Weyl points in other bands. For $k_{z}$-values between 0.17 and 0.5 there are several Fermi surfaces sheets
	and hence the focus on the integer contribution of band 48 (which is the homo for
	the originally discussed Weyl point group ($\mathrm{G}_{3}^{48}$)) is not justified,
	especially for the peaks P$_1$ at $k_{z}\approx0.27$, P$_2$ at $k_{z}\approx0.38$ and P$_3$ at $k_{z}\approx0.43$,
	which give the largest contribution to the AHC signal
	(black curve in Fig. \ref{Figure3_Theory} and Fig. \ref{fig:resolved-Chern-homo48-homo47}.A,B). 
	
	We hence have to analyze band 47 as well.
	It turns out that there are 11 Weyl point groups, which all are at
	least partially type II. All groups seem to be associated with nodal
	lines (See Fig. \ref{fig:resolved-Chern-homo48-homo47}.C). Some contain only two or four Weyl points while the rest are six-fold, which means that all but the four-fold groups are uncompensated contributing an effective
	jump of the Chern number by $\pm 1$. The groups with less than six
	points are consistent with the numerical plaquette type results
	(green curve in Fig. \ref{fig:resolved-Chern-homo48-homo47}.B),
	which does however
	not exclude the possibility that we missed some Weyl points, which
	form compensating pairs and hence do not show in the numerical results at least in the used resolution.
	
	Groups $\mathrm{G}_{5\ldots10}^{47}$ have energies between -100 and +125 meV. 
	Fig. \ref{fig:resolved-Chern-homo48-homo47}.AB
	show the comparison between the integer Chern signals for an assumed
	constant homo 48 and 47 as well as $c(k_z,E_F=45.3\mathrm{meV})$ and the resulting AHC signal.
	Groups $\mathrm{G}_{1\ldots4}^{47}$ have $k_{z}<0.15$
	and hence lay in the region where band 48 is the homo. Consequently,
	their Weyl points are fully occupied and the integer Chern signal
	is not realized if a Fermi energy is included.
	
\begin{figure}[htbp]
		\begin{centering}
			\includegraphics[width=1.0\columnwidth]{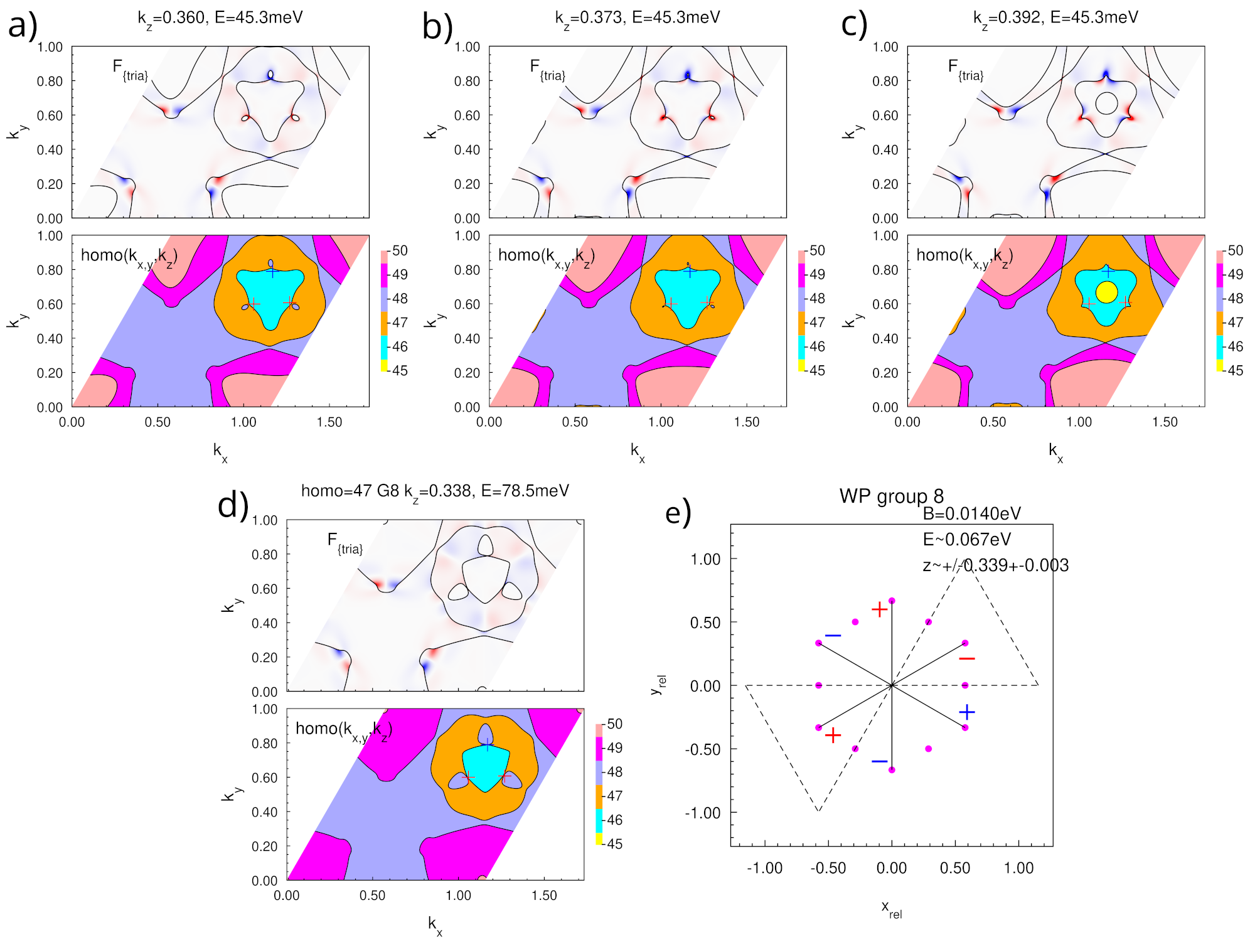}
		\end{centering}
		\caption{\label{fig:peak4_kz=00003D350}\textbf{a}-\textbf{c}: $c\left(k_{x,y},k_{z}\right)$
			(upper panel) and $\mathrm{homo}\left(k_{x,y},k_{z}\right)$ (lower panel) for
			the peak P$_2$ at Fermi energy $45.3$meV for three
			$k_{z}$ values. \textbf{d}: Same data as in \textbf{a}-\textbf{c} at $k_{z}$ and $E$ of the sixfold
			group $\mathrm{G}_{8}^{47}$. \textbf{e}: The arrangement of the Weyl points
			of $\mathrm{G}_{8}^{47}$. The Weyl points are marked with pluses, their chirality with red(positive) 
			or blue(negative) color.}
		
	\end{figure}

	Since, the groups mostly contain type II Weyl points the simple argument
	of energy independence of the AHC signal (for an ideal Weyl Hamiltonian)
	does probably hold to a lesser extend. The inevitable associated Fermi pockets lead
	to a quick deviation from the ideal case. It is therefore better to
	explain the large peaks P$_{1,2,3}$ of $c\left(k_{z},E_{\mathrm{F}}=45.3\mathrm{meV}\right)$
	via close by Weyl points. We managed to attribute peak P$_2$ at $k_{z}\approx0.38$
	(and probably also peak P$_3$ at $k_{z}\approx0.43$) to the sixfold
	group $\mathrm{G}_{8}^{47}$ of band 47, which has Weyl point energies $+66\ldots+91$
	meV. Fig. \ref{fig:peak4_kz=00003D350}.A-C show the contribution
	$c\left(k_{x,y},k_{z}\right)$ to the Chern signal $c\left(k_{z}\right)$
	for the chosen Fermi energy of $45.3$meV for three $k_{z}$-values
	in the upper panels and the highest occupied bands as function of $k_{x,y}$ 
	for this Fermi level in the lower panel. The type II Weyl points of $\mathrm{G}_{8}^{47}$
	are marked by red (positive chirality) and blue (negative chirality) pluses.
	
	Fig. \ref{fig:peak4_kz=00003D350}.D shows corresponding data at the 
	approximate Weyl point position $k_{z}=0.338$ and energy $78.5$meV of $\mathrm{G}_{8}^{47}$.
	The Weyl points sit where the pockets of band 48 touch the border between bands 46
	and 47, which is a signature of a type II Weyl point. At these parameters
	no big Berry curvature can be seen. If we lower the energy to $45.3$meV
	and shift $k_{z}$ to $0.36$ (Fig. \ref{fig:peak4_kz=00003D350}.A) the
	pockets of band 48 get smaller and have nearly vanished for $k_{Z}=0.373$
	(close to the maximum of peak P$_2$), while at the same time Berry curvature gets induced at this
	very spot. For illustration the positions of the Weyl points of $\mathrm{G}_{8}^{47}$
	are depicted in Fig. \ref{fig:peak4_kz=00003D350}.E in the first BZ. They
	all lay close to a mirror plane (solid lines) and nodal lines.
	
\begin{figure}[htbp]
		\begin{centering}
			\includegraphics[width=0.95\columnwidth]{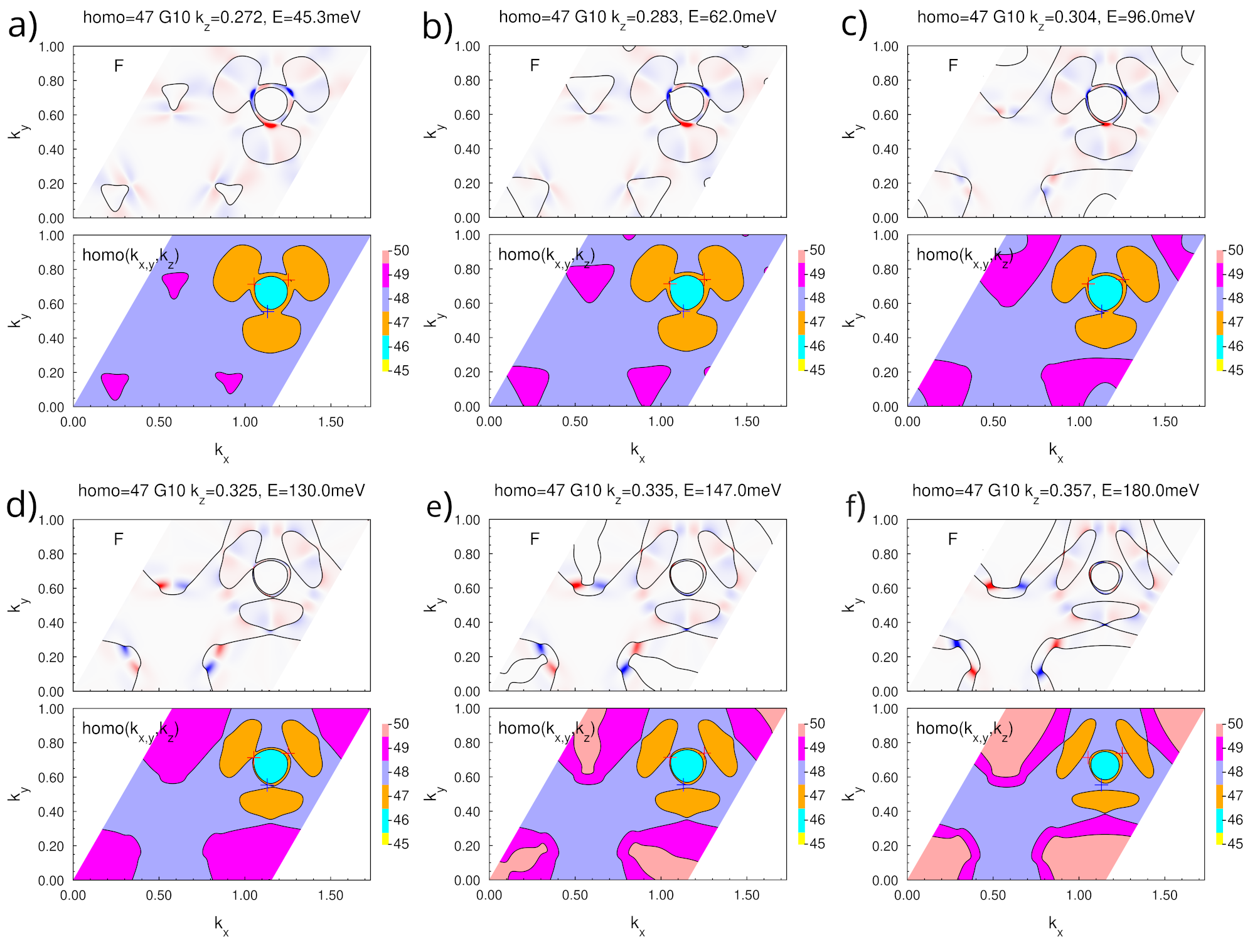}
			\par\end{centering}
		\caption{\label{fig:Peak-R-1}Peak P$_{1}$. \textbf{a}: $c\left(k_{x,y},k_{z}\right)$
			and homo structure for $k_{z}=0.272$ and $E_{F}=45.3\mathrm{meV}$
			(maximum of P$_{1}$). \textbf{b}$\ldots$\textbf{i}: evolution of
			$c\left(\boldsymbol{k}\right)$ along a linear path in $k_{z}$ and
			$E$ from P$_{1}$ (a) through the approximate WP $k_{z}$- and $E$-position
			of $G_{10}^{47}$ (d) up to a position on the other side of the WP
			position (f) , with chirality indicated by red(positive) and blue(negative)
			color resp.}
	\end{figure}

	Apparently, Weyl points need not be close in $k_{z}$-$E$-position
	to where the Berry curvature acquires a sizable signal. For peak P$_{1}$
	we can establish a (linear) path through $k_{z}$-$E$-space from
	the position of the maximum of P$_{1}$ (Fig. \ref{fig:Peak-R-1}.A) to the
	approximate Weyl point position (Fig. \ref{fig:Peak-R-1}.G) of $G_{10}^{47}$
	($k_{z}\approx0.325$, $E=130\mathrm{meV}$), along which the topology
	of the Fermi surface sheets changes little, which makes the attribution
	of the Berry curvature of peak P$_{1}$ to group $G_{10}^{47}$ likely.
	Since the WPs of $G_{10}^{47}$ are not sitting at the exact same
	$k_{z}$ and energy the Lifshitz transitions of the associated pockets
	(type II Weyl points) are not happening simultaneously along the path.
	Nevertheless, the transitions are clearly discernable. Note, that
	the WP chirality is opposite to the sign of the Bery curvature maxima.
	This is is a consequence of the jump of $\chi$ at the WP position
	as a function of $k_{z}$. Depending on which side of the WP one takes
	a cut the sign of the Berry curvature pattern switches. It should
	also be noted that the Berry curvature towards the end of the path
	(Fig. \ref{fig:Peak-R-1}.I), has the same Berry curvature sign pattern
	at the WPs as the WP's chirality, however with sizable Berry curvature
	in the triangular area close to $k_{x,y}=0$. The symmetry of the
	curvature in this area seems to approximately follow the full $\mathcal{C}_{3v}$
	symmetry (as opposed to the curvature at the WPs) and hence largely
	integrates to a small total signal.
	
\begin{figure}[htbp]
		\begin{centering}
			\includegraphics[width=0.95\columnwidth]{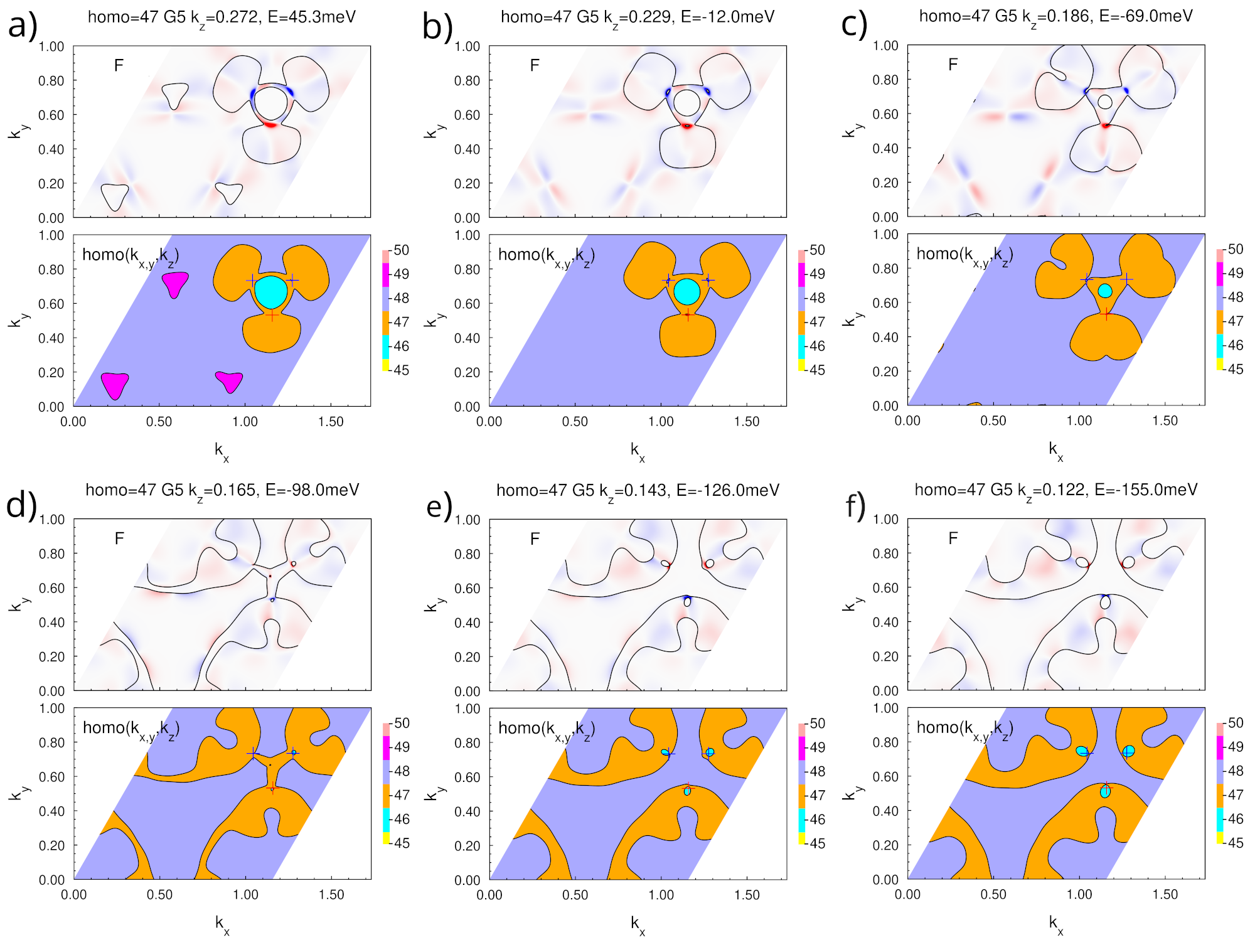}
			\par\end{centering}
		\caption{\label{fig:Peak-R-G5-1}Peak P$_{1}$. \textbf{a}: $c\left(k_{x,y},k_{z}\right)$
			and homo structure for $k_{z}=0.272$ and $E_{F}=45.3\mathrm{meV}$
			(maximum of peak P$_{1}$). \textbf{b}$\ldots$\textbf{i}: evolution
			of $c\left(\boldsymbol{k}\right)$ along a linear path in $k_{z}$
			and $E$ from peak P$_{1}$ (a) through the approximate WP $k_{z}$-
			and $E$-position of $G_{5}^{47}$ (d) up to a position on the other
			side of the WP position (f). Pluses mark the position of the WPs,
			with chirality indicated by red(positive) and blue(negative) color,
			resp.}
	\end{figure}

	It turns out that $G_{5}^{47}$ has Weyl point position at very similar
	$k_{x,y}$ as $G_{10}^{47}$ but at $k_{z}\approx0.168$ and $E\approx-98\mathrm{meV}$.
	However, in this case the Fermi surface sheets evolve into a completely
	different topology (Fig. \ref{fig:Peak-R-G5-1}). Again the Berry
	curvature pattern changes sign when going through the WPs, only this
	time the chirality pattern coincides with the BC pattern at peak P$_{1}$,
	which is due to the fact that $G_{5}^{47}$ has an effective positive
	jump of the integer Chern number. Indeed the pockets between bands
	46 and 47 are the positions which yield the BC maxima. 
	
	\subsubsection{Lattice structure}
	
	We used space group 157 (P31m) with lattice parameters $a=b=6.57316\text{Å}$
	and $c=6.16189\text{Å}$. The Wyckoff positions are: Pt at $\left(0.2619,0,0.004\right)$,
	Bi$_{1}$ at $\left(0,0,-0.359\right)$, Bi$_{2}$ at $\left(\frac{2}{3},\frac{1}{3}-0.204\right)$
	and Bi$_{3}$ at $\left(-0.3856,0,0.271\right)$.

	\subsubsection{TNL conversion into Weyl Nodes: Model Hamiltonian}

\begin{figure}[htbp]
		\begin{centering}
			\includegraphics[width=1.0\columnwidth]{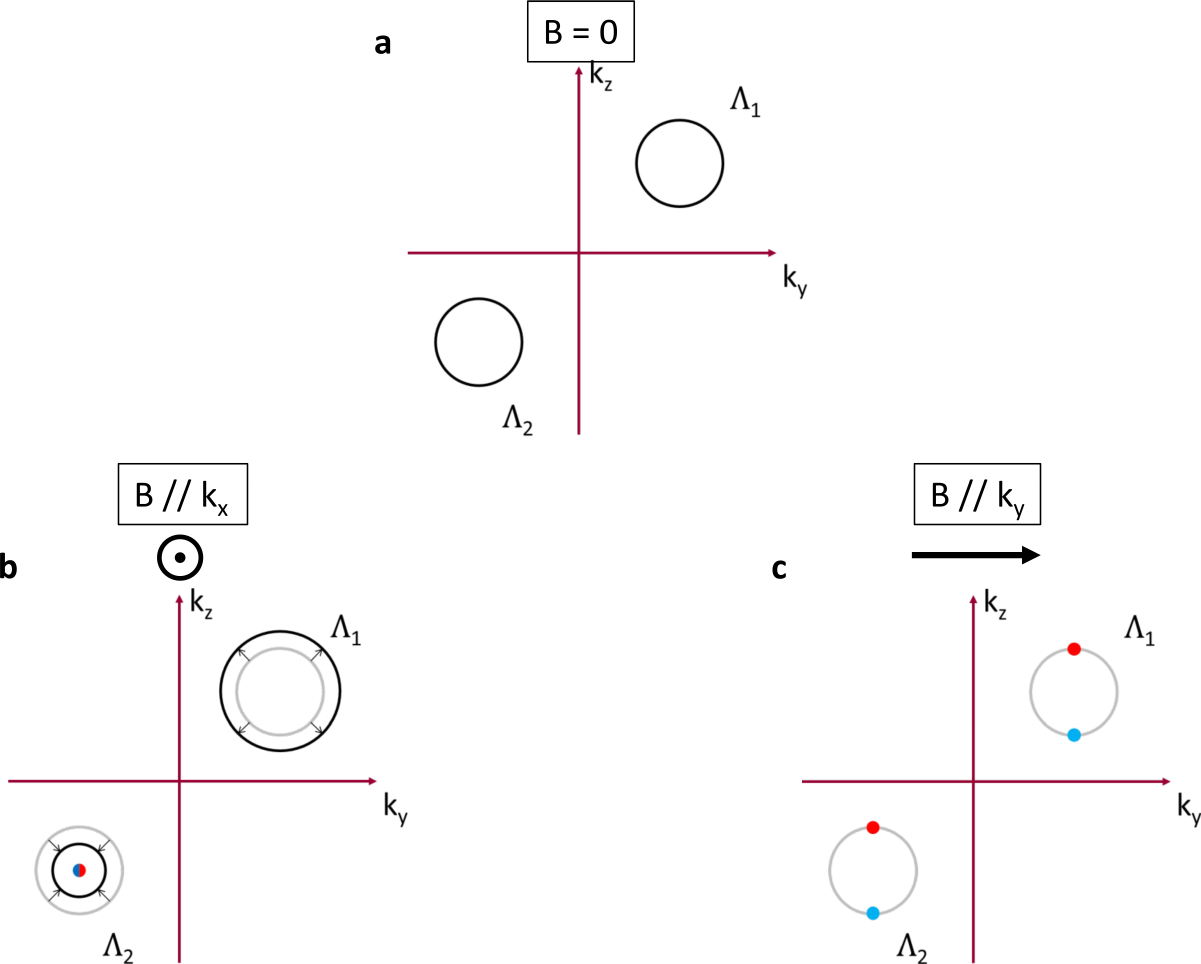}
		\end{centering}
		\caption{\label{SI fig:TNL_toy_model} Toy model showing the position of the Weyl nodes (originating from the TNL) as a function of a magnetic field which either  preserves the ${\mathcal M}_x$ mirror symmetry (a), or breaks that mirror symmetry (b,c). In each graph, the positions of the two Weyl nodes of opposite chiralities are shown in blue and red, respectively. All WNs always sit in the $k_y = 0$ plane.
			(a) shows the $k_x$ position of the two WNs when the field is along $k_x$. In this case, both WNs are always at $k_z = 0$. It can be seen that the two WNs originate from the same point at $k_x = 0$ when the field reaches the critical value. 
			(b,c) show the $k_x$ and $k_z$ positions, respectively, of the two WNs when the field is along $k_y$. As soon as $B > 0$, two WNs appear at $k_x = 0$ and $k_z = \pm 1$, and start drifting in magnetic field. This shows the non-local nature of the WN creation from TNLs.
		}
	\end{figure}

	Our starting point is a ${\bf k \cdot p}$ model for two nodal loops centered around two valleys $\Lambda_{1,2}$ related by time-reversal symmetry and located on a mirror invariant plane, which, as in the main part of the manuscript, we take as $k_x \equiv 0$. The corresponding vertical mirror symmetry can be represented as ${\mathcal M}_x = - i \sigma_x  \otimes \tau_0$ with the Pauli matrix vectors ${\boldsymbol \sigma}$ and ${\boldsymbol \tau}$ that act in spin and valley space respectively. Since time-reversal symmetry exchanges the two valleys, it can be represented as $\Theta=-i \sigma_y \otimes \tau_x$. We next introduce a two band model Hamiltonian, which generalizes the model considered in Ref.\cite{Sun2018}, that respects time-reversal and mirror symmetry and  reads: 
	\begin{equation}
		{\mathcal H}_{\bf k \cdot p}= \alpha k_x k_y \sigma_{z} \otimes \tau_z + \alpha k_x k_z \sigma_y \otimes \tau_z + \beta \left(k_0^2 + k_x^2 - k_y^2 - k_z^2 \right) \sigma_x \otimes \tau_z. 
	\end{equation}
	For $k_x \equiv 0$ two circular nodal loops of radius $k_0$  are obtained. We next consider the effect of a planar magnetic field introducing a Zeeman coupling 
	\begin{equation}
		{\mathcal H}_{Zeeman}= B \cos{\theta}~ \sigma_x \otimes \tau_0 + B \sin{\theta}~ \sigma_y \otimes \tau_0    
	\end{equation}
	where $\theta$ indicates the angle of the planar magnetic field from the $\hat{x}$ direction. It is instructive to first consider the effect of a magnetic field with $\theta \equiv 0$, in which case the vertical mirror symmetry is preserved. For small enough (positive) values of the magnetic field, one finds that the nodal line in one valley expands -- its radius is renormalized by the magnetic field to $\sqrt{k_0^2+B}$ --  while the nodal line in the opposite valley get shrunk to a radius $\sqrt{k_0^2 -B}$. At the critical value of magnetic field $B_c \equiv k_0^2$ the nodal line shrinks to a single point (see \autoref{SI fig:TNL_toy_model}.a). It is at this critical magnetic field that the conversion of the nodal line into Weyl points occurs. For $B>B_c$ zero energy states are in fact obtained at $k_y \equiv k_z \equiv 0$ and at the two mirror symmetry related momenta $k_x \equiv \pm \sqrt{B - k_0^2}$ (see \autoref{SI fig:TNL_toy_model}.a). The conversion by a mirror symmetry preserving planar magnetic field is therefore ``local" in momentum space: a nodal line shrinks to a single point out of which two Weyl nodes of opposite chirality emerge.\\
	The conversion of a nodal line into Weyl nodes is instead completely different in nature when considering a planar magnetic field that breaks the protecting vertical mirror symmetry. It in fact occurs even for infinitesimal $B$ values. Let us take the planar magnetic field to be parallel to the vertical mirror plane and therefore oriented along the $\hat{y}$ direction. We thus set $\theta=\pi/2$ in the Zeemann Hamiltonian above. 
	Using that the Hamiltonian is diagonal in valley space, 
	the eigenenergies can be then written as 
	\begin{equation}
		E^2=\alpha^2 k_x^2 k_y^2 + \left(\alpha k_x k_z \pm B \right)^2 + \beta^2 \left(k_0^2 + k_x^2 - k_y^2 - k_z^2 \right)^2
	\end{equation}
	where $\pm$ distinguishes the two valleys. 
	The spectrum is symmetric under $E \rightarrow -E$. Therefore, Weyl nodes, if present, will appear as zero-energy states. These are determined by requiring that the three squared polynomials in the equation above must simultaneously vanish. From the first term on the r.h.s. of the equation above we immediately have that the Weyl nodes will sit on the $k_y \equiv 0$ plane. The two constraints $\alpha k_x k_z \pm B \equiv 0$ and $k_0^2 + k_x^2 - k_z^2 \equiv 0$ determine the complete position of two Weyl nodes per valley. We note that since the combined ${\mathcal M}_x^{\prime}$ point group symmetry is preserved the Weyl nodes in opposite valleys are paired up to have opposite $k_z$ values and equal $k_x$ values.
	This results are sketched in \autoref{SI fig:TNL_toy_model}.c. For $B \rightarrow 0$ we find that the Weyl nodes appear at $k_x \simeq 0$ and $k_z \simeq \pm k_0$. In other words, Weyl nodes emerge out of a nodal loop with a $k_z$ distance that equals the nodal loop radius. This highlights the ``non-local" conversion of nodal loops into Weyl nodes. 

	\subsubsection{TNL gap in magnetic field and thermal activation}
	
\begin{figure}[htbp]
		\begin{centering}
			\includegraphics[width=1.0\columnwidth]{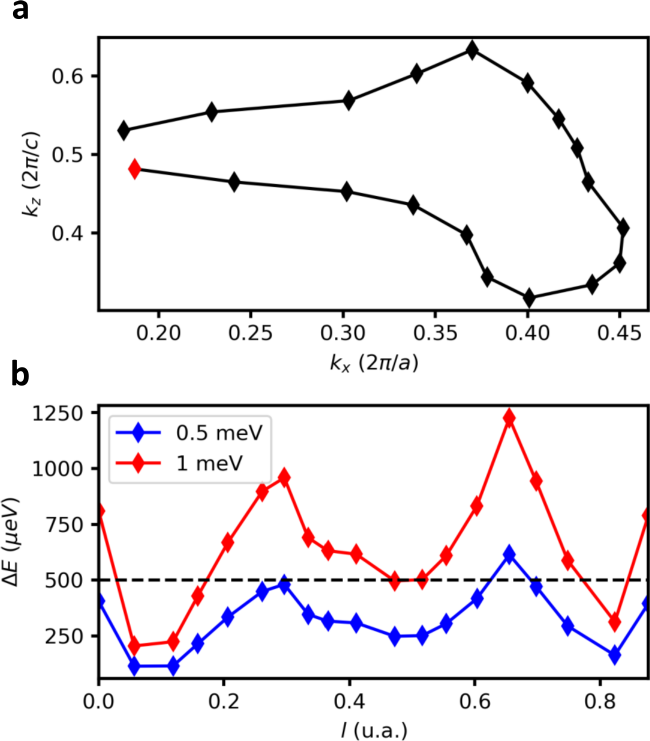}
		\end{centering}
		\caption{\label{SI fig:TNL_gap} DFT calculation of the gap opening in field at different points along the TNL.
			(a) Black diamonds: ($k_x$,$k_z$) positions of the different points considered along the TNL for the analysis (in the mirror plane $\mathcal{M}_x$); Red diamond: First point considered when going along the TNL.   
			(b) Size of the gap along the TNL, for two Zeeman energies (0.5~meV in blue, and 1~meV in red). The parameter $l$ corresponds to the length along the path going through all points in (a), starting from the red diamond.
		}
	\end{figure}

	It is possible to estimate the size of the gap at the onset field $B_{ons} = 2.8$~T by considering that is equates the thermal energy: $-g \mu_B  m  B_{ons} = k_B T$ (with $m = \pm 1/2$ for electrons). For this, we must assume a value of the g-factor. With a g-factor of 6 for band 28 (from \cite{Xing2020}), we get $-g  \mu_B m  B_{ons} \sim 486~\mu$eV, which is very consistent with $k_B T \sim 430~\mu$eV at $T = 5~$K. \\
	We can compare this to the TNL gap from DFT calculations performed for Zeeman energies $E_Z = 500~\mu$eV and $E_Z = 1$meV, although some important precautions must me taken when comparing the two, as we will detail later.
	\autoref{SI fig:TNL_gap} shows the points considered along the TNL in (a) and the gap along the TNL at both Zeeman energies in (b). The TNL is already gaped by about $500~\mu$eV in some places at $E_Z = 500~\mu$eV, while at $E_Z = 1$meV almost the entire TNL is gaped by more than $500~\mu$eV. This is consistent with the small calculation shown above.\\
	This DFT calculation cannot however be directly compared to the real system for two reasons: First, the points shown are in the mirror plane $\mathcal{M}_x$), while the WNs do not lie in the plane (they move in the $k_x$ direction under the magnetic field). Second, the TNL does not sit exactly at $E_F$, nor is it at a constant energy, meaning the temperature does not have the same effect in all parts of the TNL.
	
\end{document}